\documentclass[a4paper,11pt]{article}

\pdfoutput=1

\usepackage{amsmath,amssymb,mathtools}
\usepackage{color}
\usepackage{graphicx}
\usepackage{subfigure}
\usepackage{cite}
\usepackage[colorlinks=true,linkcolor=blue, citecolor=red, urlcolor=blue, bookmarks]{hyperref}
\usepackage{multirow,makecell}
\usepackage{textcomp}
\usepackage{wasysym}

\usepackage{tikz}

\usepackage[text={17.1cm,24.6cm},centering]{geometry} 

\numberwithin{equation}{section}

\def \be {\begin{equation}}
\def \ee {\end{equation}}
\def \ba {\begin{array}}
\def \ea {\end{array}}
\def \bea{\begin{eqnarray}}
\def \eea{\end{eqnarray}}
\def \nn {\nonumber}

\def \a {\alpha}

\def \g {\gamma}

\def \G {\Gamma}
\def \d {\delta}
\def \D {\Delta}
\def \e {\epsilon}
\def \ve {\varepsilon}
\def \m {\mu}
\def \n {\nu}

\def \l {\lambda}

\def \s {\sigma}

\def \r {\rho}

\def \th {\theta}

\def \vph {\varphi}

\def \t {\tau}

\def \cO {\mathcal O}
\def \cP {\mathcal P}
\def \cQ {\mathcal Q}
\def \cR {\mathcal R}

\def \rC {\mathrm C}

\def \rZ {\mathrm Z}

\def \p {\partial}

\def \f {\frac}

\def \tf {\tfrac}

\def \lt {\left}
\def \rt {\right}

\def \lra {\leftrightarrow}
\def \sr {\sqrt}

\def \inf {\infty}

\def \lag {\langle}
\def \rag {\rangle}

\def \dd {\mathrm{d}}
\def \ep {\mathrm{e}}
\def \ii {\mathrm{i}}

\def \tr {\textrm{tr}}

\def \and {{~\textrm{and}~}}
\def \with {{~\textrm{with}~}}

\begin{document}

\title{\textbf{Subsystem distance after a local operator quench}}
\author{
Jiaju Zhang$^{1}$ 
and
Pasquale Calabrese$^{1,2}$ 
}

\maketitle
 \vspace{-10mm}
\begin{center}
{\it
$^{1}$SISSA and INFN, Via Bonomea 265, 34136 Trieste, Italy\\\vspace{1mm}
$^{2}$International Centre for Theoretical Physics (ICTP), Strada Costiera 11, 34151 Trieste, Italy
}
\vspace{10mm}
\end{center}

\begin{abstract}

We investigate the time evolution of the subsystem trace distance and Schatten distances after local operator quenches in two-dimensional conformal field theory (CFT) and in one-dimensional quantum spin chains. We focus on the case of a subsystem being an interval embedded in the infinite line. The initial state is prepared by inserting a local operator in the ground state of the theory. We only consider the cases in which the inserted local operator is a primary field or a sum of several primaries. While a nonchiral primary operator can excite both left-moving and right-moving quasiparticles, a holomorphic primary operator only excites a right-moving quasiparticle and an anti-holomorphic primary operator only excites a left-moving one. The reduced density matrix (RDM) of an interval hosting a quasiparticle is orthogonal to the RDM of the interval without any quasiparticles. Moreover, the RDMs of two intervals hosting quasiparticles at different positions are also orthogonal to each other. We calculate numerically the entanglement entropy, R\'enyi entropy, trace distance, and Schatten distances in time-dependent states excited by different local operators in the critical Ising and XX spin chains. These results match the CFT predictions in the proper limit.

\end{abstract}

\baselineskip 18pt
\thispagestyle{empty}
\newpage


\tableofcontents

\section{Introduction}

In the nearly past two decades, the investigation of entanglement entropy have greatly deepened our understanding of quantum many-body systems, quantum field theories,
and gravity \cite{Amico:2007ag,calabrese2009entanglement,Laflorencie:2015eck,Ryu:2006bv}. 
Besides the entanglement entropy, it is also important to quantify the difference of the reduced density matrices (RDMs) of a subsystem in different states of the total system, especially in non-equilibrium configurations.
One quantitative description of the difference between RDMs is through a distance;
however there are many different distances \cite{nielsen2010quantum,watrous2018theory}, but, among these, the trace distance is special \cite{fagotti2013reduced}.
Recently, the subsystem trace distance in low-lying states of two-dimensional (2D) conformal field theories (CFTs) have been calculated in
Refs. \cite{Zhang:2019wqo,Zhang:2019itb}.
In this paper, we consider the subsystem trace distance (and more generally also other Schatten distances) in a special kind of non-equilibrium state known as local operator
quench \cite{Nozaki:2014hna,Nozaki:2014uaa,He:2014mwa}.
We focus on (1+1)-dimensional CFTs and on spin chains whose continuum low-energy limit is described by them.

A quantum quench is the rapid evolution of an initial state that is not an energy eigenstate.
In a CFT, there are two main families of quantum quenches \cite{cc-16}:
in a global quench, the initial state is globally different from the energy eigenstates, as e.g. in \cite{Calabrese:2005in,Calabrese:2006rx};
in a local quench, the initial state is locally different from an energy eigenstate, as for examples in the joining and splitting
quenches \cite{Calabrese:2007mtj,sd-11,Shimaji:2018czt} or in the local operator quench \cite{Nozaki:2014hna,Nozaki:2014uaa,He:2014mwa}.
In this paper, we consider the subsystem trace distance in latter category of quenches:
at an initial time $t=0$, a local operator is inserted somewhere in the ground state of the theory, and then the system is let evolve without further disturbance.
The time evolution of entanglement and R\'enyi entropies, as well as of local operators, has been intensively studied in the literature
\cite{kkm-08,Nozaki:2014hna,Nozaki:2014uaa,He:2014mwa,cmt-14,csst-14,dch-15,gh-15,cgh-15,nnm-15,cv-15,Caputa:2016yzn,nw-16,dkk-16,ckt-17,nw-17,kt-17,
Bhattacharyya:2019ifi,Caputa:2019avh,Kusuki:2019avm}.
The main goal of this paper is to investigate {\it the distance between the RDMs at different times after a local operator quench}.
While in previous studies, the focus has been on primary or descendent operators \cite{He:2014mwa,cgh-15},
here we also consider the case in which the local operator is a sum of several primary fields
(as we shall see, this is not only done for academic curiosity, but because it is the right setup to describe the local operator quenches in spin chains and
lattice models).
This is different from the case of several operators inserted simultaneously at different positions investigated in \cite{Caputa:2019avh,Kusuki:2019avm}.
For two-dimensional CFTs, we get analytical results for the time evolution of the entanglement entropy, R\'enyi entropies, trace distance, and Schatten distances.

We also study the same quantities in one-dimensional quantum spin chains whose continuum low-energy limit is described by a CFT.
Specifically, we consider the critical Ising spin chain (whose continuum limit is the 2D free massless fermion theory, which is a 2D CFT with central charge $c=1/2$),
and the XX chain (2D free massless compact boson theory with central charge $c=1$).
The entanglement and R\'enyi entropies after some local operator quenches in the Ising spin chain have been already investigated in \cite{Caputa:2016yzn}.
In this paper, we will calculate numerically the entanglement entropy, the R\'enyi entropies, the trace distance, and the Schatten distances in the critical Ising spin chain,
as well as in the XX spin chain.
We get numerical results that match the analytical CFT results in the proper limit.

The remaining part of the paper is arranged as follows.
In section~\ref{secCFT} we first review the 2D CFT approach to the entanglement and R\'enyi entropies in the local operator quench \cite{He:2014mwa} and then
calculate the trace and Schatten distances.
In section~\ref{sec:sc} we  review the construction of the reduced density matrix in XY spin chains and then
calculate numerically the entanglement entropy, the R\'enyi entropy, the trace distance, and the Schatten distances.
In section~\ref{secInside} we discuss some generalizations of our findings and, in particular, we present explicit results for the case
of the operation insertion inside the subsystem of interest.
In section~\ref{secCon} we conclude with discussions.
In appendix~\ref{appIde}, we identify the local operators in 2D CFT with those of critical spin chains for both the critical Ising and XX spin chains.

{\bf Note added}:
During the final stage of preparation of this manuscript, Ref. \cite{Bhattacharyya:2019ifi} appeared where a similar, but in fact different, local operator quench was investigated.
The states considered in \cite{Bhattacharyya:2019ifi} are superpositions of  density matrices and the total system is generally in a mixed state.
The states we consider are superpositions of operators and the total system is always in a pure state.
The entanglement and R\'enyi entropies in \cite{Bhattacharyya:2019ifi} and in this paper behave similarly.
Besides the entanglement and R\'enyi entropies, we also calculate the subsystem trace and Schatten distances.

\section{Operator quenches in two-dimensional CFT}\label{secCFT}

We review the 2D CFT approach to the quench by the insertion of local operator \cite{Nozaki:2014hna,Nozaki:2014uaa,He:2014mwa},
and generalize it to the case in which the local operator is a sum of several primary fields.
We calculate R\'enyi entropies as well as the trace and Schatten distances of the RDMs at different times after the local operator quench.

\subsection{The protocol}

We consider a 2D CFT on an infinite line.
At $t=0$ a local operator $\cO$ is inserted at $x=\ell_0$.
The normalized density matrix of whole system after the quench is
\be
\r_\cO(t) = \f{\ep^{-\ii H t} \ep^{-\e H} \cO(\ell_0) |0\rag \lag0| \cO^\dag(\ell_0) \ep^{-\e H} \ep^{\ii H t}}
     {\lag0| \cO^\dag(\ell_0) \ep^{-2\e H} \cO(\ell_0) |0\rag},
     \label{rhot}
\ee
with $\e$ being the UV regularization \cite{Calabrese:2005in} of the locally excited state.
Universal effects manifest in the limit of time $t$ and distances much larger than $\epsilon$.
As the theory is a CFT, one can equivalently consider the limit $\e\to 0$ and all other scales finite.
The Euclidean spacetime is the complex plane with coordinates
\be \label{wbarw}
w= \t-\ii x, ~~ \bar w= \t+\ii x ,
\ee
where $\t=\ii t$ is the Euclidean time (we work in units in which the speed of the light is $v=1$).
In the present convention, the holomorphic operators are right-moving modes and the anti-holomorphic operators are left moving modes.
The density matrix (\ref{rhot}) may be written as
\be \label{rcOt}
\r_\cO(t) = \f{\cO(w_2,\bar w_2) |0\rag\lag0| \cO^\dag(w_1,\bar w_1)}{\lag \cO^\dag(w_1,\bar w_1) \cO(w_2,\bar w_2) \rag_{\rC}},
\ee
with operators inserted at the positions
\bea
&& w_1 = \e-\ii(t+\ell_0) , ~~ \bar w_1 = \e-\ii(t-\ell_0), \nn\\
&& w_2 = -\e-\ii(t+\ell_0), ~~ \bar w_2 = -\e-\ii(t-\ell_0).
\label{points}
\eea
Notice that $\bar w_i$ are not the complex conjugates of $w_i$: indeed, in principle, one should first perform the calculation for imaginary time $\tau$ being real and, only
at the end of the calculation, perform the analytic continuation to imaginary time \cite{Calabrese:2006rx,cc-07}:
this is equivalent to work directly with the positions in Eq. \eqref{points}.
When the local operator $\cO$ is primary with conformal dimension $(h_\cO,\bar h_\cO)$, i.e. with scaling dimension $\D_\cO=h_\cO+\bar h_\cO$ and spin $s_\cO=h_\cO-\bar h_\cO$, the normalization of the density matrix is
\be \label{normalization}
\lag \cO^\dag(w_1,\bar w_1) \cO(w_2,\bar w_2) \rag_{\rC} = \f{\a_\cO}{(2\e)^{2\D_\cO}}.
\ee
Note that, independently of the normalization amplitude $\a_\cO$, the density matrix is always properly normalized so that $\tr\r_\cO(t)=1$.

\subsection{The entanglement entropies}

We consider the subsystem $A=[0,\ell]$ and the RDM $\r_{A,\cO}(t)=\tr_{\bar A}\r_\cO(t)$, with $\bar A $ being the complement of $A$.
Without loss of generality, we choose the interval to be on the right of the inserted operator and we fix $\ell_0=-\ell$
(with a conformal transformation we can get an arbitrary value of $\ell_0$, see e.g. \cite{Nozaki:2014hna,Nozaki:2014uaa}, thus we only consider $\ell_0=-\ell$).
At $t=0$, the degrees of freedom within the interval $A$ are not yet affected by the inserted operator, and so the RDM equals the RDM in the ground state
$ \r_{A,0}\equiv\r_{A,\cO}(0)$. Consequently, the initial R\'enyi entropy is just the ground state R\'enyi entropy $S_{A,0}^{(n)}\equiv S_A^{(n)}(0) $.
At $t>0$ the increase of the R\'enyi entropy due to the insertion of the local operator can be written as \cite{Calabrese:2004eu,cc-09,Nozaki:2014hna,Nozaki:2014uaa}
\bea \label{DSAntequivCorre}
&& \D S_A^{(n)}(t) \equiv S_A^{(n)}(t) - S_A^{(n)}(0)
          = -\f{1}{n-1} \log \f{\tr\r^n_{A,\cO}(t)}{\tr\r^n_{A,0}} \nn\\
&& \phantom{\D S_A^{(n)}(t)}
        = -\f{1}{n-1} \log
          \f{\lag \cO^\dag(w_1,\bar w_1) \cO(w_2,\bar w_2) \cdots 
           \cO(w_{2n},\bar w_{2n}) \rag_{\rC_n}}
           {\lag \cO^\dag(w_1,\bar w_1) \cO(w_2,\bar w_2) \rag_{\rC}^n}.
\eea
In the numerator of (\ref{DSAntequivCorre}), the operators $\cO^\dag(w_{2j-1},\bar w_{2j-1})$, $\cO(w_{2j},\bar w_{2j})$ with $j=1,2,\cdots,n$ are inserted on the
$j$-th replica of the $n$-fold plane $\rC_n$. The $2n$-point correlation function on $\rC_n$ are calculated by mapping it to complex plane with coordinate $z$ by
means of the conformal transformation \cite{Calabrese:2004eu,cc-09}
\be
z = \Big( \f{w}{w+\ii\ell} \Big)^{1/n}, ~~
\bar z = \Big( \f{\bar w}{\bar w-\ii\ell} \Big)^{1/n}.
\ee
We denote by $(z_1,\bar z_1),(z_2,\bar z_2),\cdots,(z_{2n},\bar z_{2n})$ the location of the operator insertion in the $z$ plane.
In real time, when $0<t<\ell$ and $t>2\ell$, we have
\be
z_{2j-1} = \ep^{\f{2\pi\ii j}{n}} \Big( \f{\ell-t-\ii\e}{2\ell-t-\ii\e} \Big)^{1/n}, ~~
z_{2j} = \ep^{\f{2\pi\ii j}{n}} \Big( \f{\ell-t+\ii\e}{2\ell-t+\ii\e} \Big)^{1/n},
\ee
while for $\ell<t<2\ell$
\be
z_{2j-1} = \ep^{\f{2\pi\ii}{n}(j-\f12)} \Big( \f{t-\ell+\ii\e}{2\ell-t-\ii\e} \Big)^{1/n}, ~~
z_{2j} = \ep^{\f{2\pi\ii}{n}(j+\f12)} \Big( \f{t-\ell-\ii\e}{2\ell-t+\ii\e} \Big)^{1/n}.
\ee
At any time $t>0$, for the anti-holomorphic we have
\be
\bar z_{2j-1} = \ep^{-\f{2\pi\ii j}{n}} \Big( \f{\ell+t+\ii\e}{2\ell+t+\ii\e} \Big)^{1/n}, ~~
\bar z_{2j} = \ep^{-\f{2\pi\ii j}{n}} \Big( \f{\ell+t-\ii\e}{2\ell+t-\ii\e} \Big)^{1/n}.
\ee
The $2n$-point function in Eq. \eqref{DSAntequivCorre} can be expanded in conformal blocks \cite{bpz-84}.
In the limit $\e\to 0$ (equivalently $\ell,t\gg \e$), one block is much larger than all the others.
From the above expressions for $z_k$ and $\bar z_k$, the dominant channel for the holomorphic sector for $0<t<\ell$ and $t>2\ell$ is
\be
(z_1,z_2)(z_3,z_4)\cdots(z_{2n-1},z_{2n}),
\ee
which is the same for the anti-holomorphic sector at any time.
Conversely, in the holomorphic sector for $\ell<t<2\ell$, the dominant channel is
\be
(z_2,z_3)(z_4,z_5)\cdots(z_{2n},z_{1}).
\label{ch2}
\ee
%
Using the fact that the conformal block for the non-chiral primary operator $\cO$ in the channel \eqref{ch2} is simply related to the {\it quantum dimension} $d_\cO$ of $\cO$,
the time evolution of the R\'enyi entropy (for $t,\ell\gg \e$) is very easily worked out as \cite{He:2014mwa}
\be
\D S_A^{(n)}(t) =
\lt\{\ba{ll}
 0     & 0<t<\ell {~\rm and~} t>2\ell \\
 \log d_\cO & \ell<t<2\ell
\ea\rt..
\label{DS}
\ee
Note that $\D S_A^{(n)}(t)$ is independent of $n$.
The result \eqref{DS} can be interpreted in terms of the entangled quasiparticle picture of Ref. \cite{Calabrese:2005in}.
At $t=0$ a pair of entangled quasiparticles, one left-moving and the other right-moving, are excited at the position where the operator is inserted $x=-\ell$;
at $t=\ell$ the right-moving quasiparticle enters in the interval from the left boundary at $x=0$;
at $t=2\ell$ the right-moving quasiparticle leaves the interval from the right boundary at $x=\ell$.

When $\cO$ is a purely holomorphic primary operator $\cP$ or a purely anti-holomorphic primary operator $\cQ$, the R\'enyi entropy is constant in time, i.e.
\be
\D S_A^{(n)}(t) = 0.
\label{DSP}
\ee

\subsection{The distances}

Given the time dependent RDM $\r_A(t)=\r_{A,\cO}(t)$, we define a normalized Schatten distance with $n\geq1$
\be
D_A^{(n)}(t_1,t_2) = \Big( \f{\tr | \r_A(t_1)-\r_A(t_2) |^n}{2\tr\r_{A,0}^n} \Big)^{1/n},
\label{Dndef}
\ee
where, again, $\r_{A,0}$ is the RDM of the ground state and $\r_A(0)=\r_{A,0}$.
Notice that the true Schatten distance is given by $D_A^{(n)}(t_1,t_2)(\tr\r_{A,0}^n)^{1/n}$, but we introduced the normalization factor $(\tr\r_{A,0}^n)^{1/n}$ for convenience
in the field theoretical calculations (see, e.g., \cite{Zhang:2019itb});
with some abuse of notation we will refer to $D_A^{(n)}(t_1,t_2)$ as Schatten distances, rarely specifying they are normalized as in \eqref{Dndef}.

Following the approach of Refs. \cite{Zhang:2019wqo,Zhang:2019itb}, we first calculate the Schatten distance for general even integer $n_e$
\be \label{DAnet1t2}
D_A^{(n_e)}(t_1,t_2) = \Big( \f{\tr ( \r_A(t_1)-\r_A(t_2) )^{n_e}}{2\tr\r_{A,0}^{n_e}} \Big)^{1/{n_e}},
\ee
and then take the analytical continuation arbitrary real $n_e\to n$ to get the Schatten distance.
In particular, for $n_e\to1$ we get trace distance
\be
D_A(t_1,t_2) = \lim_{n_e\to1} D_A^{(n_e)}(t_1,t_2)=\f12 \tr | \r_A(t_1) - \r_A(t_2) |.
\ee

We evaluate (\ref{DAnet1t2}) using replica trick. The crucial point is the limit $\e\to 0$. Indeed, in this limit,
we can use once again the conformal block expansion and isolate the dominant term(s).
When $\cO$ is a non-chiral primary operator or a purely holomorphic operator $\cP$, for $0<t<\ell$ and $t>2\ell$ we can straightforwardly show that
\be \label{utile1}
\f{\tr ( \r_{A,\cO}^m(t)\r_{A,0}^{n-m} )}{\tr\r_{A,0}^n} = 1 ~\Rightarrow~ \r_{A,\cO}(t) = \r_{A,0},
\ee
and at $\ell<t<2\ell$
\be \label{utile2}
\f{\tr ( \r_{A,\cO}^m(t)\r_{A,0}^{n-m} )}{\tr\r_{A,0}^n} = 0 ~\Rightarrow~ \r_{A,\cO}(t) \perp \r_{A,0}.
\ee
Moreover for $\ell<t_1<2\ell$, $\ell<t_2<2\ell$ and $t_1 \neq t_2$
\be \label{utile3}
\f{\tr ( \r_{A,\cO}^m(t_1)\r_{A,\cO}^{n-m}(t_2) )}{\tr\r_{A,0}^n} = 0 ~\Rightarrow~ \r_{A,\cO}(t_1) \perp \r_{A,\cO}(t_2).
\ee
When $\cO$ is a purely anti-holomorphic primary operator ${\cQ}$, we have at any time $t>0$
\be \label{utile4}
\f{\tr ( \r_{A,{\cQ}}^m(t)\r_{A,0}^{n-m} )}{\tr\r_{A,0}^n} = 1 ~\Rightarrow~ \r_{A,{\cQ}}(t) = \r_{A,0}.
\ee
Eqs. (\ref{utile1}), (\ref{utile2}), (\ref{utile3}) and (\ref{utile4}) holds for arbitrary integers $1 \leq m \leq n-1$ and they are also robust under any permutations of the RDMs.
Consequently, the RDMs at different times can only be either the same or orthogonal.
(This is reminiscent of the famous Anderson orthogonality theorem \cite{anderson1967infrared,anderson1967ground}.)
Furthermore, all these results can be simply interpreted in the quasiparticles picture.
The insertion of a non-chiral operator excite both left- and right-moving quasiparticles, while a purely holomorphic operator only excites a right-moving quasiparticle and a purely anti-holomorphic operator only excites a left-moving quasiparticle.
When there is no quasiparticle in the interval, the RDM just equals the RDM of the ground state.
The RDM of the interval hosting a quasiparticle is not only orthogonal to the RDM of the interval without any quasiparticle but is also orthogonal to the RDM of the interval
hosting a quasiparticle at a different position.

\begin{figure}[t]
 \centering
 \includegraphics[height=0.24\textwidth]{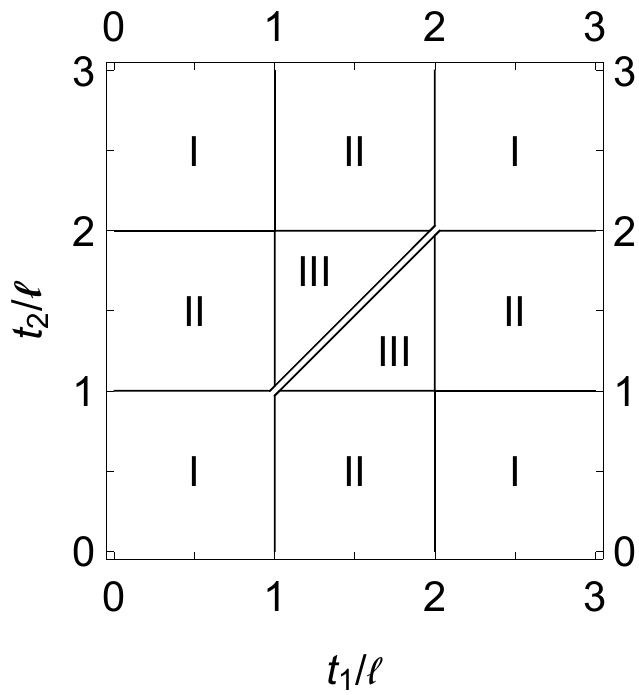}\\
 \caption{The division of the values of $t_1,t_2$ into three regions according to the position of the quasiparticle.}\label{div}
\end{figure}

The above equations are all we need to calculate the Schatten distances of even order.
In the $(t_1,t_2)$ plane, we identify three regions with different behavior of the distances, that we denote as regions I, II, and III, and are shown in figure~\ref{div}.
Note that the narrow region between the two parts of region III belongs to region I.

When $\cO$ is a non-chiral primary operator,
in region I, $\rho_A(t_1)$ equals $\rho_A(t_2)$ and hence all distances vanish.
In region II and III they are orthogonal, hence $\tr (\rho_A(t_1)-\rho_A(t_2))^{n_e}= \tr \rho_A^{n_e}(t_1)+\tr \rho^{n_e}_A(t_2)$, because all mixed terms like
 $\tr (\rho_A^m(t_1) \rho_A^{n_e-m}(t_2))$ vanish (notice that for odd $n_o$ we would get $\tr (\rho_A(t_1)-\rho_A(t_2))^{n_o}= \tr \rho_A^{n_o}(t_1)-\tr \rho^{n_o}_A(t_2)$).
Using the explicit values for $\tr \rho_A^n(t)$ in Eq. \eqref{DS}, we get the normalized Schatten distances (which represent the analytic continuation to arbitrary $n$)
\be
D_A^{(n)}(t_1,t_2)=
\lt\{\ba{ll}
 0                     & {\rm region~I} \\
 \Big( \f{1+1/d_\cO^{n-1}}{2} \Big)^{1/n} & {\rm region~II} \\
 \f{1}{d_\cO^{(n-1)/n}}           & {\rm region~III}
\ea\rt.,
\label{DnO}
\ee
that lead to the trace distance
\be
D_A(t_1,t_2)=
\lt\{\ba{ll}
 0 & {\rm region~I} \\
 1 & {\rm regions~II~and~III}
\ea\rt..
\ee
Eq. \eqref{DnO} clearly shows why the normalization \eqref{Dndef} is very convenient in these field theoretical calculations.
The relative entropy of two orthogonal density matrices is divergent and not well-defined.
Indeed, the calculation of $S(\r_A(t)\|\r_{A,0})$ using replica trick of \cite{Lashkari:2014yva,Lashkari:2015dia} leads to divergence.
When $\cO$ is a purely holomorphic primary operator $\cP$, repeating the calculation above but with the R\'enyi entropy in \eqref{DSP},
we get the Schatten distance and trace distance
\be
D_A^{(n)}(t_1,t_2)=D_A(t_1,t_2)=
\lt\{\ba{ll}
 0 & {\rm region~I} \\
 1 & {\rm regions~II~and~III}
\ea\rt..
\ee
When $\cO$ is a purely anti-holomorphic primary operator $\cQ$, we obtain
\be
D_A^{(n)}(t_1,t_2)=D_A(t_1,t_2)= 0 {\rm~~regions~I,~II~and~III}.
\ee

\subsection{Excitation as sum of primary fields}
\label{sumpr}

We now consider the case when the inserted operator is the sum of several primary operators.
To sum several primary operators with different scaling dimensions, we need to introduce coefficients with dimensions of length.
If these dimensions are finite, in the limit of small UV cutoff $\e \to 0$, the primary operators with the largest scaling dimensions give the largest contribution to
R\'enyi entropies and Schatten distances.
Thus, we only need to consider the case when the primary operators have the same scaling dimensions.
The most general operator of this kind has the form
\be \label{cOcPcQcR}
\cO(w,\bar w) = \mu \cP(w) + \nu \cQ(\bar w) + \sum_i \xi_i \cR_i(w,\bar w),
\ee
where the operators $\cP(w)$, $\cQ(\bar w)$, $\cR_i(w,\bar w)$ are orthonormal primary operators with the same scaling dimension $\D_\cO>0$,
with $\cP$ is purely holomorphic, $\cQ$ purely anti-holomorphic, and $\cR_i$ non-chiral.
For the operator $\cO$, we define the ratios of purely holomorphic part, purely anti-holomorphic part, and each non-chiral part as
\bea\label{pqri}
&& p = \f{|\m|^2}{|\m|^2+|\n|^2+\sum_i|\xi_i|^2}, \nn\\
&& q = \f{|\n|^2}{|\m|^2+|\n|^2+\sum_i|\xi_i|^2}, \nn\\
&& r_i = \f{|\xi_i|^2}{|\m|^2+|\n|^2+\sum_j|\xi_j|^2}.
\eea
These ratios are in the range $[0,1]$ and satisfy
\be
p+q+\sum_i r_i=1.
\ee
The normalized RDM of the interval $A$ after the insertion of $\cO$ is
\be
\r_{A,\cO}(t) = p \r_{A,\cP}(t) + q \r_{A,0} + \sum_i r_i \r_{A,\cR_i}(t) + \cdots,
\ee
with $\cdots$ denoting cross terms that will not contribute to our calculations because of orthogonality in the limit $\e\to 0$.
We have used that $\r_{A,\cQ}(t)=\r_{A,0}$ as $\cQ$ is purely anti-holomorphic.
Following exactly the same logic as in the previous section and using in particular the orthogonality of the RDMs for $\ell<t<2\ell$,
we have that $\tr \rho_A^n(t)$ becomes, in the limit $\e\to0$, the sum of three independent pieces
($\tr \rho_A^n(t)= p^n \tr (\r_{A,\cP}(t))^n +q^n \tr \r_{A,0}^n + \sum_i r_i^n \tr (\r_{A,\cR_i}(t))^n $).
Using then Eqs. \eqref{DS} and \eqref{DSP},
 we simply get the R\'enyi entropy after the local operator quench of $\cO$
\be
\D S_A^{(n)}(t) =
\lt\{\ba{ll}
 0 & 0<t<\ell {~\rm and~} t>2\ell \\
 -\f{1}{n-1} \log \Big( p^n + q^n + \sum_i \f{r_i^n}{d_{\cR_i}^{n-1}} \Big) & \ell<t<2\ell
\ea\rt.,
\label{DS1}
\ee
and the entanglement entropy
\be
\D S_A(t) =
\lt\{\ba{ll}
 0               & 0<t<\ell {~\rm and~} t>2\ell \\
 -p\log p - q \log q - \sum_i r_i \log\f{r_i}{d_{\cR_i}} & \ell<t<2\ell
\ea\rt.,
\label{DSS}
\ee
with $d_{\cR_i}$ being the quantum dimension of the non-chiral primary operator $\cR_i$.

In a very similar way, we also get the Schatten distances
\be
D_A^{(n)}(t_1,t_2)=
\lt\{\ba{ll}
 0                     & {\rm region~I} \\
 \Big[ \f12 \Big( p^n + (1-q)^n + \sum_i \f{r_i^n}{d_{\cR_i}^{n-1}} \Big) \Big]^{1/n} & {\rm region~II} \\
 \Big( p^n + \sum_i \f{r_i^n}{d_{\cR_i}^{n-1}} \Big)^{1/n}           & {\rm region~III}
\ea\rt.,
\label{DDn}
\ee
and the trace distance
\be
D_A(t_1,t_2)=
\lt\{\ba{ll}
 0  & {\rm region~I} \\
 1-q & {\rm regions~II~and~III}
\ea\rt..
\label{DD1}
\ee
It is remarkable that the trace distance only depends on $q$, the ratio of the anti-holomorphic part.

The ratios (\ref{pqri}) have a clear physical interpretation. The above results of entropies and distances indicate that the insertion of a purely holomorphic operator only excites a right-moving quasiparticle, a purely anti-holomorphic operator only excites a left-moving quasiparticle, and the insertion of a nonchiral operator excites a pair of right-moving and left-moving quasiparticles with entanglement that depends on the quantum dimension of the inserted operator. For the insertion of the operator (\ref{cOcPcQcR}), the ratios (\ref{pqri}) are probabilities of various quasiparticles being excited. The ratio $p$ is the probability that only a right-moving quasiparticle is excited, $q$ is the probability that only a left-moving quasiparticle is excited, and $r_i$ is the probability that a pair of special type of right-moving and left-moving quasiparticles are excited.

The probabilistic interpretation of the ratios also explains why the entropies (\ref{DS1}) and (\ref{DSS}) are symmetric under the exchange $p\lra q$, while the distances (\ref{DDn}) and (\ref{DD1}) are not.
The whole system is always in a pure state, and so the entanglement entropies of the subsystem and its complement are always the same, and the right-moving and left-moving quasiparticles contribute the same to the entropies.
As the interval is located at the right of the inserted operator, the right-moving and left-moving quasiparticles contribute differently to the distances.
For example, the ratio $1-q$ in (\ref{DD1}) means the probability that a right-moving quasiparticle is excited, including the case with only a right-moving quasiparticle and the cases with both a right-moving quasiparticle and a left-moving one, i.e. that $1-q=p+\sum_i r_i$.

\begin{figure}[tbp]
 \centering
 \includegraphics[height=0.24\textwidth]{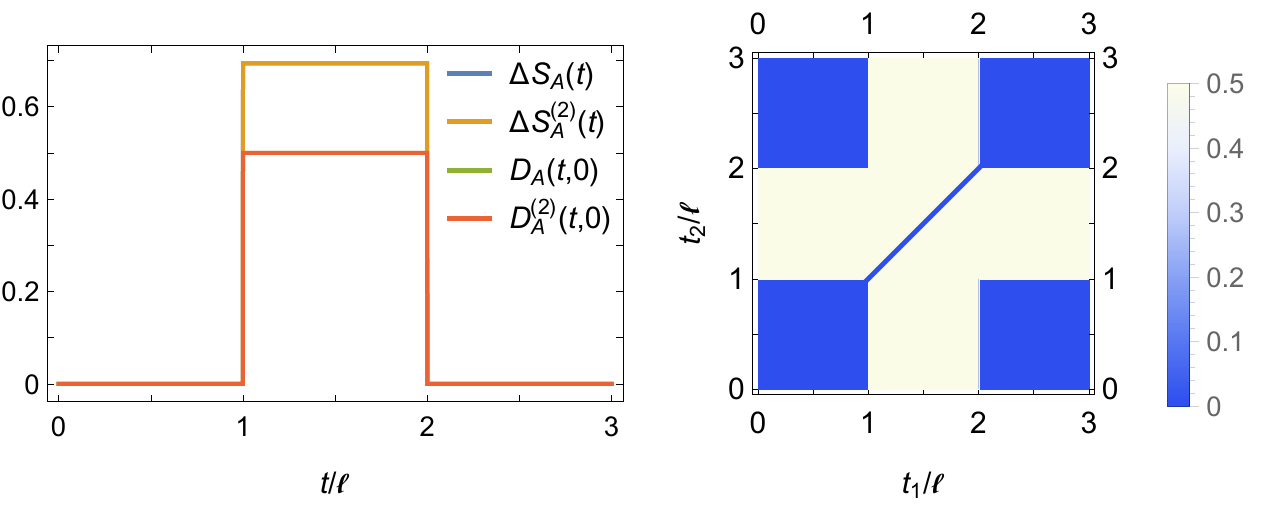}\\
 \caption{
 The quantities of interest for this paper after a local operator quench with the insertion of a sum of holomorphic and anti-holomorphic primary fields
 with equal weight (i.e. ${q}=\f12$).
 Left: the entanglement entropy $\D S_A(t)$ (\ref{DSAt}), the second R\'enyi entropy $\D S^{(2)}_A(t)$ (\ref{DSAnt}),
 the trace distance $D_A(t,0)$ (\ref{DAnt1t2Dt1t2}), and the second Schatten distance $D^{(2)}_A(t,0)$ (\ref{DAnt1t2Dt1t2}).
 Note that the entanglement and R\'enyi entropies coincide (this is an accident for $q=1/2$) and also the trace distance and the second Schatten distance
 (this is instead always true).
 Right: Trace distance $D_A(t_1,t_2)$ (\ref{DAnt1t2Dt1t2}) in the entire $(t_1,t_2)$ plane.
 For $q=1/2$, the trace and Schatten distances can only be zero or $1/2$.}\label{CFTEES2TDD2}
\end{figure}

One special case of (\ref{cOcPcQcR}), of relevance for the following, is when $\cO$ is the sum of a purely holomorphic and a purely anti-holomorphic primary operators
\be \label{cOwbarw}
\cO(w,\bar w) = \m \cP(w) + \n \cQ(\bar w).
\ee
Now the ratio of the anti-holomorphic part is
\be \label{ratio}
q = \f{|\n|^2}{|\m|^2+|\n|^2},
\ee
and the ratio of the holomorphic part is $p=1-q$.
The limit for $r_i\to 0$ of Eqs. (\ref{DS1}-\ref{DD1}) provides
\be \label{DSAnt}
\D S_A^{(n)}(t) =
\lt\{\ba{ll}
 0                               & 0<t<\ell {~\rm and~} t>2\ell \\
 -\f{1}{n-1} \log \big[ q^n + (1-q)^n \big] & \ell<t<2\ell
\ea\rt.,
\ee
\be \label{DSAt}
\D S_A(t) =
\lt\{\ba{ll}
 0               & 0<t<\ell {~\rm and~} t>2\ell \\
 -q\log q - (1-q) \log (1-q) & \ell<t<2\ell
\ea\rt.,
\ee
\be \label{DAnt1t2Dt1t2}
D_A^{(n)}(t_1,t_2)=D_A(t_1,t_2)=
\lt\{\ba{ll}
 0  & {\rm region~I} \\
 1-q & {\rm regions~II~and~III}
\ea\rt..
\ee
These CFT results (\ref{DSAnt}), (\ref{DSAt}) and (\ref{DAnt1t2Dt1t2}) apply to all the states we will consider in the critical spin chains in this paper.
An example for $q=1/2$ is plotted in figure~\ref{CFTEES2TDD2}.

\subsection{The effect of the cutoff in CFT}

\begin{figure}[t]
 \centering
 \includegraphics[height=0.24\textwidth]{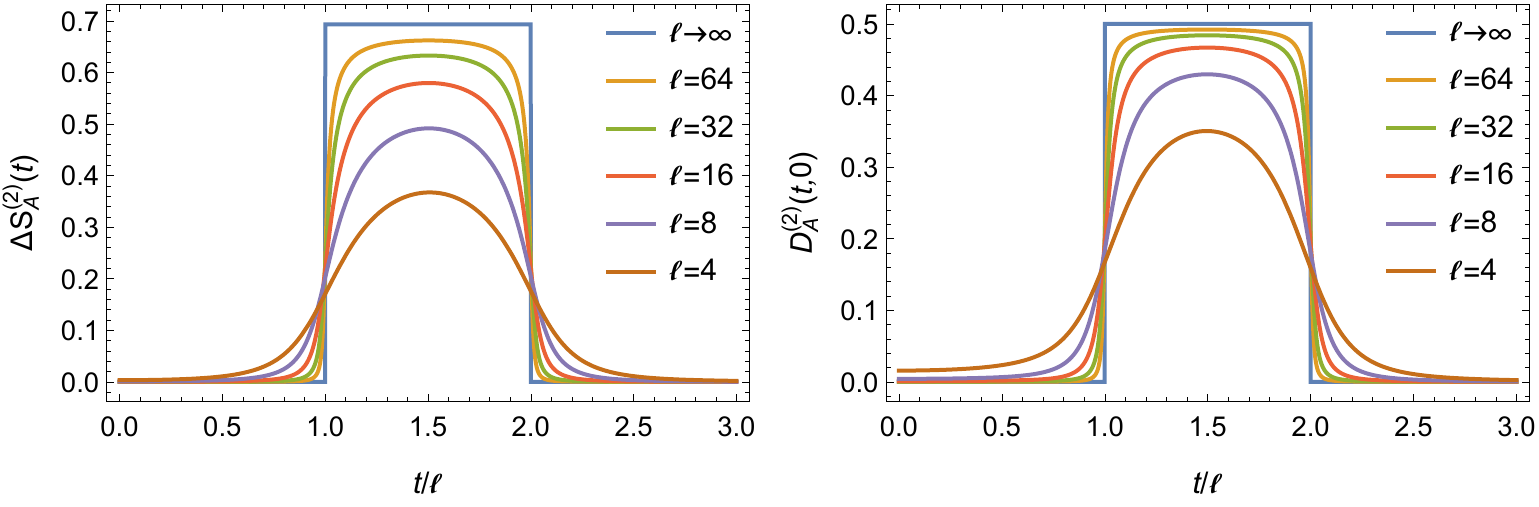}\\
 \caption{Effect of the cutoff in a two-dimensional CFT. Here we consider the free massless fermion theory and the insertion of the operator (\ref{esempio0}).
 We set the UV cutoff of the operator to be $\e=1$ and we report results for various values of $\ell$.
 The time evolution of the second order R\'enyi entropy $\D S_A^{(2)}(t)$ (\ref{esempio1}) and Schatten 2-distance $D_A^{(2)}(t,0)$ (\ref{esempio2})
 are reported on the left and right respectively. They slowly approach the results for $\e\to0$ as $\ell\to\infty$.
 }\label{CFTfinitecutoff}
\end{figure}

The lattice spacing in spin chains provides a natural cutoff both for the entire theory and for the inserted local operator.
Thus, to recover the CFT prediction, we will need to take the interval length $\ell$ to be much larger than the lattice spacing.
It is rather natural to wonder about the effect of the cutoff directly in CFT to compare, at least qualitatively, with the lattice results.
These effects are not universal, so as an example, we work in the free (Majorana) fermion theory and we consider the insertion of the operator
\be \label{esempio0}
\cO(w,\bar w) = \f{1}{\sr 2} ( \psi(w) + \bar\psi(\bar w) ),
\ee
with the ratio of the anti-holomorphic part $q=\f12$.
Note that $\psi(w)$ and $\bar\psi(\bar w)$ are primary operators with conformal weights $(\f12,0)$ and $(0,\f12)$, respectively, while $\cO(w,\bar w)$ is neither primary nor descendant.
Using the previous prescriptions we get the density matrix of the time-dependent state for the whole system
\be
\r_{\cO}(t) = \e ( \psi(w_2) + \bar\psi(\bar w_2) ) |0\rag \lag0| ( \psi(w_1) + \bar\psi(\bar w_1) ),
\ee
from which we deduce the RDM $\r_{A,\cO}(t)=\tr_{\bar A}\r_{\cO}(t)$.
For general $n$, the explicit results for a finite cutoff $\e$ may be obtained using known results for the $2n$-point functions of Majorana fermions
(see, e.g., \cite{DiFrancesco:1997nk}), but they are long and not very illuminating.
We only report the concrete example of $n=2$ for which we get
\bea
&& \f{\tr\r_{A,\cO}^2(t)}{\tr\r_{A,0}^2} =
\e^2 \lag ( \psi(w_1) + \bar\psi(\bar w_1) ) ( \psi(w_2) + \bar\psi(\bar w_2) ) ( \psi(w_3) + \bar\psi(\bar w_3) ) ( \psi(w_4) + \bar\psi(\bar w_4) ) \rag_{\rC_2} \nn\\
&& \phantom{\f{\tr\r_{A,\cO}^2(t)}{\tr\r_{A,0}^2}}
=\ell^2\e^2
\Big[
-\f{1}{w_1w_2(w_1+\ii \ell)(w_2+\ii \ell)} \f{z_1^2z_2^2}{(z_1^2-z_2^2)^2}
-\f{1}{\bar w_1\bar w_2(\bar w_1-\ii \ell)(\bar w_2-\ii \ell)} \f{\bar z_1^2\bar z_2^2}{(\bar z_1^2-\bar z_2^2)^2} \nn\\
&& \phantom{\f{\tr\r_{A,\cO}^2(t)}{\tr\r_{A,0}^2} =}
+\sr{\f{z_1z_2\bar z_1\bar z_2}{w_1w_2(w_1+\ii \ell)(w_2+\ii \ell)\bar w_1\bar w_2(\bar w_1-\ii \ell)(\bar w_2-\ii \ell)}}
 \f{z_1\bar z_1 + z_2 \bar z_2}{(z_1^2-z_2^2)(\bar z_1^2-\bar z_2^2)} \nn\\
&& \phantom{\f{\tr\r_{A,\cO}^2(t)}{\tr\r_{A,0}^2} =}
 -\f{1}{16} \Big( \f{1}{w_1(w_1+\ii \ell)} + \f{1}{\bar w_1(\bar w_1-\ii \ell)} \Big)
      \Big( \f{1}{w_2(w_2+\ii \ell)} + \f{1}{\bar w_2(\bar w_2-\ii \ell)} \Big)
\Big].
\eea
The resulting evolution of the second R\'enyi entropy increase is
\be \label{esempio1}
\D S_A^{(2)}(t) = - \log \f{\tr\r_{A,\cO}^2(t)}{\tr\r_{A,0}^2},
\ee
and it is shown in figure~\ref{CFTfinitecutoff}.
Another useful result is
\bea
&& \f{\tr(\r_{A,\cO}(t)\r_{A,0})}{\tr\r_{A,0}^2} =
\e \lag ( \psi(w_1) + \bar\psi(\bar w_1) ) ( \psi(w_2) + \bar\psi(\bar w_2) ) \rag_{\rC_2} \\
&& \phantom{\f{\tr(\r_{A,\cO}(t)\r_{A,0})}{\tr\r_{A,0}^2}}=
\f{\ii\ell\e}{2}\Big(
 \sr{\f{z_1z_2}{w_1w_2(w_1+\ii \ell)(w_2+\ii \ell)}}\f{1}{z_1 - z_2}
- \sr{\f{\bar z_1\bar z_2}{\bar w_1\bar w_2(\bar w_1-\ii \ell)(\bar w_2-\ii \ell)}} \f{1}{\bar z_1 - \bar z_2}
\Big). \nn
\eea
from which we calculate the Schatten 2-distance
\be \label{esempio2}
D_A^{(2)}(t,0) = \Big[ \f{\tr(\r_{A,\cO}(t)-\r_{A,0})^2}{2\tr\r_{A,0}^2} \Big]^{1/2},
\ee
shown also in Figure~\ref{CFTfinitecutoff}.
As $\ell\to\inf$, Eqs. \eqref{esempio1} and \eqref{esempio2} match (\ref{DSAnt}) and (\ref{DAnt1t2Dt1t2}) with ${q}=1/2$ respectively.

\section{Local operator quenches in spin chains}
\label{sec:sc}

In this section, we first review some basic features of the XY spin chain, with particular emphasis on the two special gapless cases, i.e., the critical Ising and XX models.
We focus on those aspects that will be useful to the present paper, more general details can be found in, e.g., Refs. \cite{Lieb:1961fr,Berganza:2011mh,calabrese2012quantum,Zhang:2019itb,pfeuty1970one,katsura1962statistical,f-book}
and references therein.

The Hamiltonian of the XY spin chain in transverse field is
\be \label{HXY}
H = - \sum_{j=1}^L \Big( \f{1+\g}{4}\s_j^x\s_{j+1}^x + \f{1-\g}{4}\s_j^y\s_{j+1}^y + \f{\l}{2}\s_j^z \Big),
\ee
with $\s_j^{x,y,z}$ denoting the Pauli matrices and $L$ the total number of sites in the chain.
We choose periodic boundary conditions $\s_{j+L}^{x,y,z}=\s_j^{x,y,z}$ and we focus on the ground state.
We only consider $L$ being four times an integer and we will finally take the $L \to \inf$ limit.
For $\g=1$, Eq. \eqref{HXY} reduces to the Ising spin chain which is critical for $\l=1$ and
its continuum limit is the free massless fermion theory, which is a 2D CFT with central charge $c=\f12$.
For $\g=0$, Eq. \eqref{HXY} is the XX spin chain, which is critical for $| \lambda | \leq 1$ with continuum limit given by a free massless compact boson (for $| \lambda | < 1$),
which is a 2D CFT with central charge $c=1$.

The Hamiltonian of the XY spin chain \eqref{HXY} can be exactly diagonalized by applying successively the Jordan-Wigner transformation, the Fourier transformation,
and a Bogoliubov rotation \cite{Lieb:1961fr,katsura1962statistical}
\be
a_j = \Big(\prod_{i=1}^{j-1}\s_i^z\Big)\s_j^+, ~~
a_j^\dag = \Big(\prod_{i=1}^{j-1}\s_i^z\Big)\s_j^-,
\ee
\be
b_k = \f{1}{\sr{L}}\sum_{j=1}^L\ep^{\ii j \vph_k}a_j, ~~
b_k^\dag = \f{1}{\sr{L}}\sum_{j=1}^L\ep^{-\ii j \vph_k}a_j^\dag,
\ee
\be
c_k = b_k \cos\f{\th_k}{2} + \ii b_{-k}^\dag \sin\f{\th_k}{2}, ~~
c_k^\dag = b_k^\dag \cos\f{\th_k}{2} - \ii b_{-k} \sin\f{\th_k}{2}.
\ee
Here $\s_j^\pm = \f12 ( \s_j^x \pm \ii \s_j^y )$ and $\varphi_k=2\pi k/L$.
The ground state is in the Neveu-Schwarz (NS) sector with antiperiodic boundary conditions for the fermions
$a_{j+L} = - a_j, ~~ a^\dag_{j+L} = - a^\dag_j$ so that the momenta are quantized as
\be
k =-\f{L}{2}+\f12, \cdots, -\f12, \f12, \cdots, \f{L}{2}-\f12.
\ee
The Bogoliubov angle $\th_k \in [-\pi,\pi]$ is
\be
\ep^{\ii \th_k} = \f{\l - \cos\vph_k + \ii \g\sin\vph_k}{\ve_k},
\ee
and the energies of the single particle modes are
\be
\ve_k = \sr{(\l - \cos\vph_k)^2+\g^2\sin^2\vph_k}.
\label{epsk}
\ee
In diagonal form the Hamiltonian in the NS sector is
\be
H = \sum_k \ve_k \Big( c_k^\dag c_k -\f12 \Big).
\label{DH}
\ee
Since $\ve_k\geq0$, the ground state $|0\rag$ is annihilated by all the modes $c_k$
\be
c_k |0\rag=0, ~ k =-\f{L}{2}+\f12, \cdots,\f{L}{2}-\f12.
\ee

For critical Ising spin chain with $\g=\l=1$, we have
\be
\label{eps_IS}
\ve_k = 2\sin\f{\pi|k|}{L}, ~~
\ee
while for critical XX spin chain with $\g=\l=0$, we have
\be
\ve_k =\Big|\cos\f{2\pi k}{L} \Big|. ~~
\ee


\subsection{Local operator quench in XY spin chain}\label{appLOQ}

The RDM of one interval in the XY spin chain can be generically written in terms of the correlation matrix restricted to the subsystem
by means of Wick theorem \cite{chung2001density,Vidal:2002rm,peschel2003calculation,Latorre:2003kg,Alba:2009th,Calabrese:2005in}.
From these, we can easily get R\'enyi entropies and distances (indeed entanglement and R\'enyi entropies for some local operator quenches
in the Ising spin chain have been already considered in Ref. \cite{Caputa:2016yzn}).
Hence, In the following, we explicitly calculate the time dependence of the correlation matrix after a local operator quench.


It is convenient to work with the correlation function of the Majorana modes defined as
\be \label{d2jm1d2j}
d_{2j-1}=a_j + a^\dag_j, ~~
d_{2j}=\ii(a_j - a^\dag_j).
\ee
For a block of $\ell$ consecutive spins, say $A=[1,\ell]$, the $2\ell\times2\ell$ correlation matrix is
\be
\G_{m_1 m_2} = \lag d_{m_1} d_{m_2} \rag - \d_{m_1m_2},
\ee
with $m_1,m_2=1,2,\cdots,2\ell$.
The entries of the correlation matrix $\G$ are usually parametrized in terms of three functions $f,g,h$ defined as
\bea
&& \lag d_{2j_1-1} d_{2j_2-1} \rag = \d_{j_1j_2} + f_{j_1j_2}, \nn\\
&& \lag d_{2j_1} d_{2j_2} \rag = \d_{j_1j_2} + h_{j_1j_2}, \nn\\
&& \lag d_{2j_1-1} d_{2j_2} \rag = g_{j_1j_2}, \nn\\
&& \lag d_{2j_1} d_{2j_2-1} \rag = -g_{j_2j_1}.
\eea
For example, in the ground state of the XY chain, one generically has
\bea
&& f_{j_1j_2} = h_{j_1j_2} =0, ~~ g_{j_1j_2} = g_{j_2-j_1}, \nn\\
&& g_j = -\f{2\ii}{L} \sum_k \cos( j\vph_k - \th_k ). \label{gj}
\eea
In particular, in the thermodynamic limit, $L\to\inf$, we have for the critical Ising spin chain
\be
g_j = -\f{\ii}{\pi} \f{1}{j+\f12},
\ee
and for the XX spin chain in zero magnetic field
\be
g_j =
\lt\{\ba{ll}
\f{2\ii}{\pi j} \sin\f{\pi j}{2} & j \neq 0 \\
0 & j=0
\ea\rt..
\ee

Now we consider the time evolution of a state after a local operator quench in the XY spin chain (in Ref. \cite{em-09} a similar protocol has been
considered in the gapped phase of the XY spin chain).
At time $t=0$ the spin chain is in the ground state when we insert a local operator at the site $j=\ell_0$.
Let us focus on the insertion of $d_{2\ell_0-1}$ from which many other operators' insertions are easily reconstructed, as we shall see.
For $t>0$, the (normalized) time-dependent state may be written in the two equivalent forms
\bea \label{epmiiHtd2jm1}
&& \ep^{-\ii H t} d_{2\ell_0-1} |0\rag = \f{1}{L} \sum_{k,j} \ep^{-\ii(j-\ell_0)\vph_k-\ii\ve_k t} d_{2j-1} |0\rag \nn\\
&& \phantom{\ep^{-\ii H t} d_{2\ell_0-1} |0\rag} =
  \f{\ii}{L} \sum_{k,j} \ep^{-\ii[(j-\ell_0)\vph_k-\th_k]-\ii\ve_k t} d_{2j} |0\rag.
\eea
The entries of the correlation matrix $\G$ in the state (\ref{epmiiHtd2jm1}) are easily worked out using
Wick theorem and the ground state expectations values $\lag 0 | d_{2j_1-1} d_{2j_2-1} | 0 \rag=\lag 0 | d_{2j_1} d_{2j_2} | 0 \rag = \d_{j_1j_2}$;
the final results are
\bea \label{CMfhg1}
&& f_{j_1j_2} = \f{2\ii}{L^2} \sum_{k_1,k_2} \sin[ (j_1-\ell_0)\vph_{k_1} - (j_2-\ell_0)\vph_{k_2} + ( \ve_{k_1} - \ve_{k_2} )t ], \nn\\
&& h_{j_1j_2} = \f{2\ii}{L^2} \sum_{k_1,k_2} \sin[ (j_1-\ell_0)\vph_{k_1} - \th_{k_1} - (j_2-\ell_0)\vph_{k_2} + \th_{k_2} + ( \ve_{k_1} - \ve_{k_2} )t ], \nn\\
&& g_{j_1j_2} = g_{j_2-j_1} + \f{2\ii}{L^2} \sum_{k_1,k_2} \cos[ (j_1-\ell_0)\vph_{k_1} - (j_2-\ell_0)\vph_{k_2} + \th_{k_2} + ( \ve_{k_1} - \ve_{k_2} )t ],
\eea
with $g(j)$ in (\ref{gj}). In the $L \to \inf$ limit, the sum over $k$ becomes an integration as
\be
\f{1}{L}\sum_k \to \f{1}{2\pi}\int_{-\pi}^\pi \dd\vph.
\ee
Hence, the entries of the correlation matrix $\G$ in the XY spin chain in the thermodynamic limit are
\bea \label{CMfhg2}
&& f_{j_1j_2} = \f{\ii}{2\pi^2} \int_{-\pi}^\pi \!\!\dd\vph_1 \int_{-\pi}^\pi \!\!\dd\vph_2
        \sin[ (j_1-\ell_0)\vph_1 - (j_2-\ell_0)\vph_2 + ( \ve_{\vph_1} - \ve_{\vph_2} )t ], \nn\\
&& h_{j_1j_2} = \f{\ii}{2\pi^2} \int_{-\pi}^\pi \!\!\dd\vph_1 \int_{-\pi}^\pi \!\!\dd\vph_2
        \sin[ (j_1-\ell_0)\vph_1 - \th_{\vph_1} - (j_2-\ell_0)\vph_2 + \th_{\vph_2} + ( \ve_{\vph_1} - \ve_{\vph_2} )t ], \\
&& g_{j_1j_2} =g_{j_2-j_1}
       + \f{\ii}{2\pi^2} \int_{-\pi}^\pi \!\!\dd\vph_1 \int_{-\pi}^\pi \!\!\dd\vph_2
        \cos[ (j_1-\ell_0)\vph_1 - (j_2-\ell_0)\vph_2 + \th_{\vph_2} + ( \ve_{\vph_1} - \ve_{\vph_2} )t ]. \nn
\eea
These numerical integrals converge very slowly, and one should be very careful for large $\ell$.


\subsection{The critical Ising spin chain}\label{secIsing}

\begin{figure}[t]
 \centering
 \includegraphics[height=0.24\textwidth]{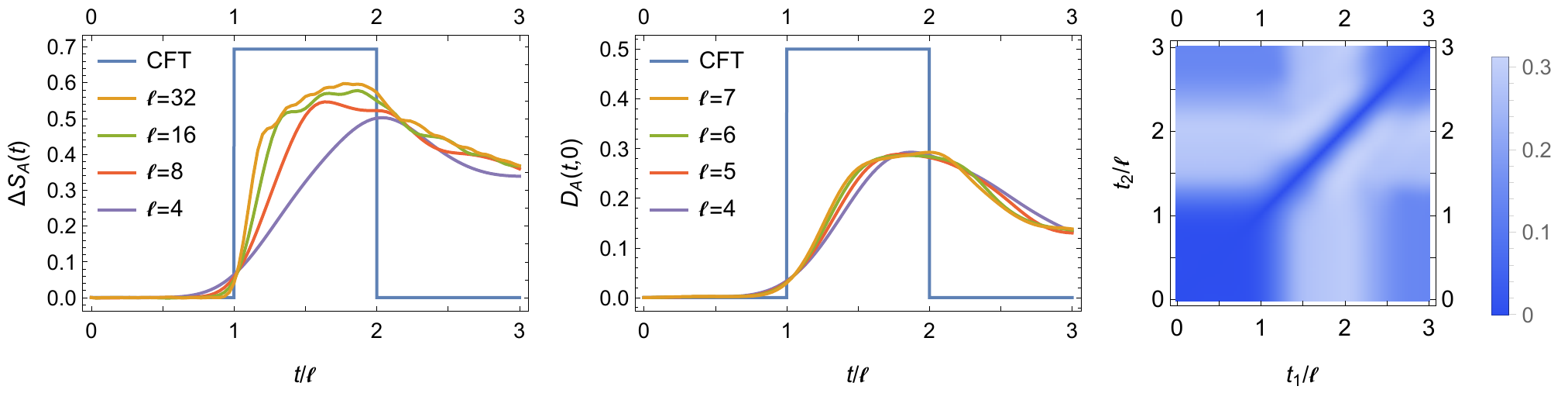}\\
%
 \includegraphics[height=0.24\textwidth]{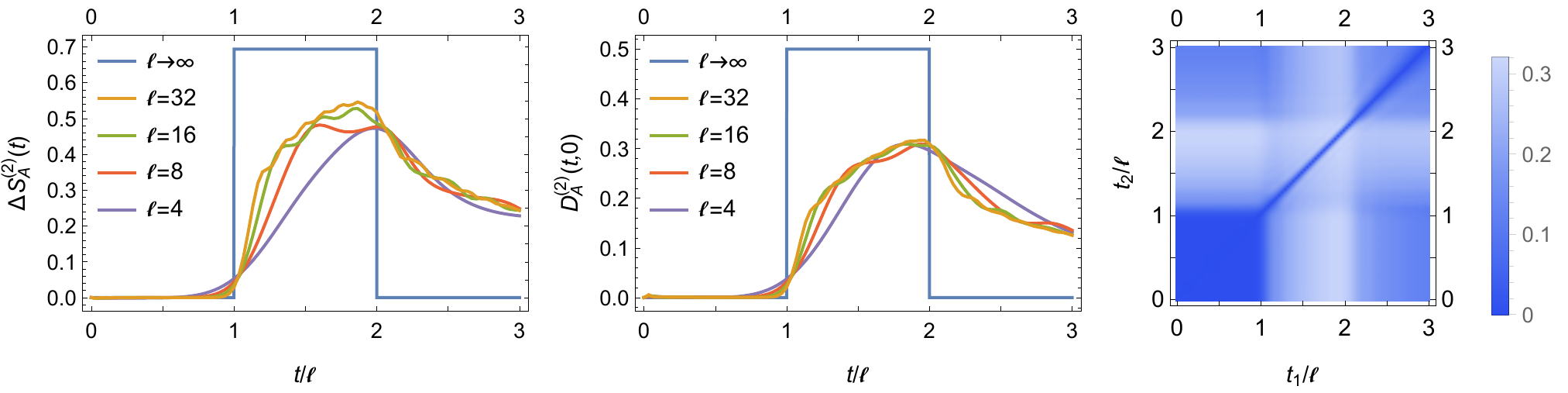}\\
 \caption{Time evolution after a local operator quench from a state with the insertion of the operator $d_{-2\ell+1}$ in the critical Ising model.
 We report entanglement and R\'enyi entropies, together with trace and Schatten distances for the spin block $A=[1,\ell]$.
 The first and second rows correspond to order one and two respectively.
 The plots in the first column are the time dependence of the entropy increase.
 The ones in the second column are the distances from the initial state for several fixed lengths $\ell$.
 The 3rd column reports colour plots of the distances between two different times at fixed $\ell$ ($\ell=7$ for the trace distance and $\ell=32$ for the Schatten one).
}
 \label{RealIsingS2D2}
\end{figure}

In this section, we present our numerical analysis of the subsystem distances after local operator quench for
the critical Ising chain (i.e. Hamiltonian (\ref{HXY}) with $\g=\l=1$).
At $t=0$, the spin chain is at the ground state $|0\rag$, and we begin by considering the insertion of the Majorana operator $d_{2j_0-1}$ at the site $\ell_0=-\ell+1$.
The time evolved state is $\ep^{-\ii H t} d_{-2\ell+1} |0\rag$ and the
corresponding density matrix is
\be
\r(t) = \ep^{-\ii H t} d_{-2\ell+1} |0\rag \lag0| d_{-2\ell+1} \ep^{\ii H t}.
\ee
The subsystem of interest is the interval $A$ consisting of the first $\ell$ sites, i.e. $A= [1,\ell]$
and the time-dependent RDM is $\r_A(t) = \tr_{\bar A}\r(t)$.

The elements of the $2\ell\times2\ell$ correlation matrix $\G$ for this local operator quench are straightforwardly written in terms of those in Eqs. (\ref{CMfhg1})
using the relation between Majorana and Dirac fermion \eqref{d2jm1d2j}.
The entanglement and R\'enyi entropies are then easily calculated in terms of the eigenvalues of this matrix using standard techniques \cite{chung2001density,Vidal:2002rm,peschel2003calculation,Latorre:2003kg}.
Also the Schatten distances for even $n$ can be obtained for large $\ell$ \cite{Zhang:2019wqo,Zhang:2019itb}, using the product rules for
Gaussian fermionic matrices \cite{Fagotti:2010yr,bb-69}.
For the trace distance instead only small subsystems (say with $\ell$ up to 7) may be worked out by explicitly constructing the entire
$2^\ell\times 2^\ell$ density matrix from the correlation one, see for details Refs. \cite{Zhang:2019wqo,Zhang:2019itb}.

\begin{figure}[t]
 \centering
 \includegraphics[height=0.24\textwidth]{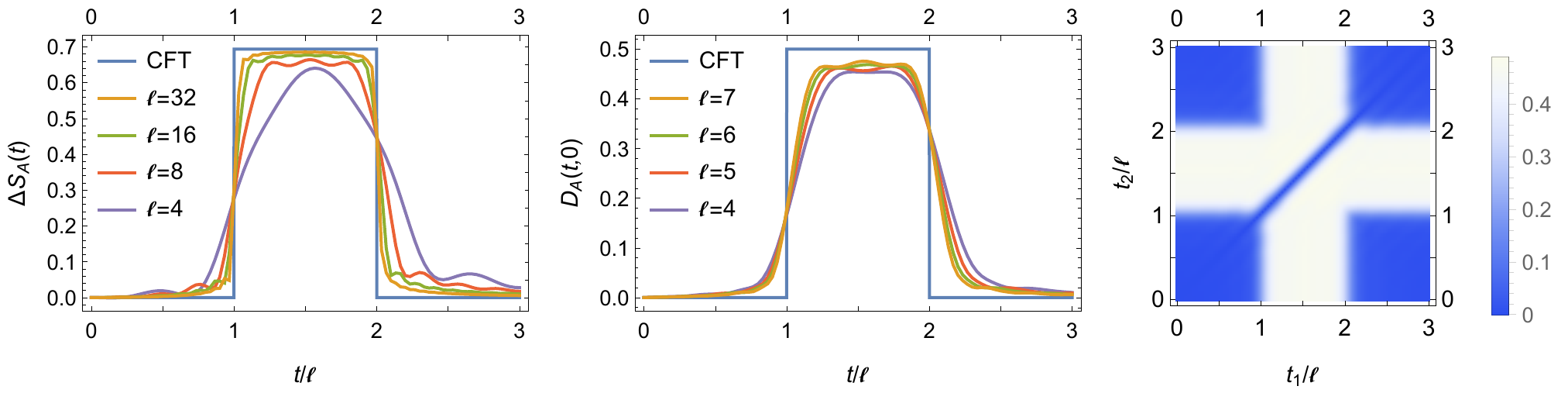}\\
%
 \includegraphics[height=0.24\textwidth]{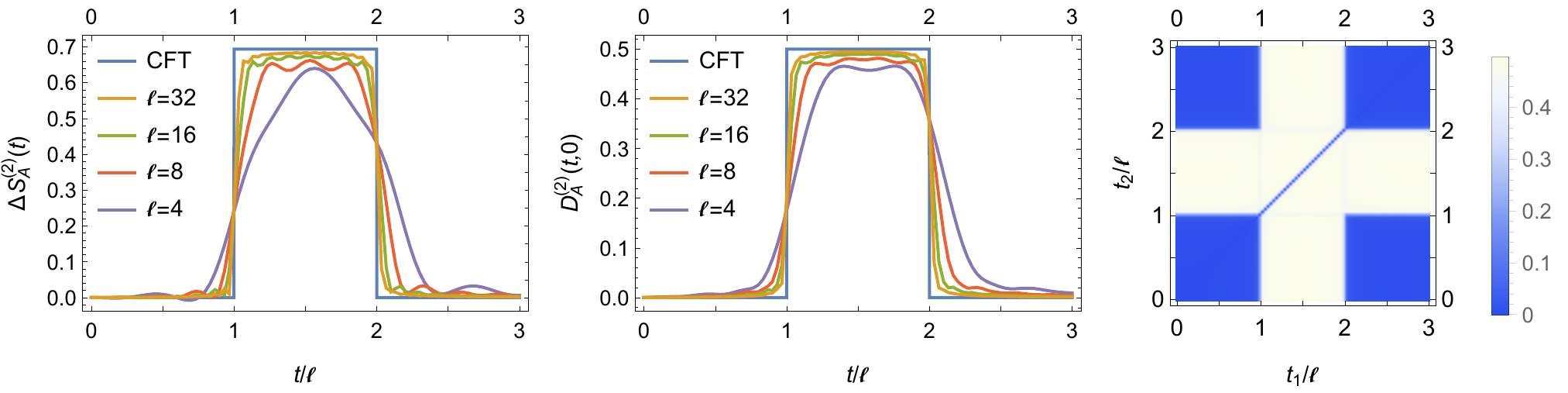}\\
 \caption{The same as in Figure \ref{RealIsingS2D2} for the {\it linearized} Ising model.
 Note the much faster approach to the CFT predictions compared to the actual spin chain in Figure \ref{RealIsingS2D2}.
 }\label{LinearizedIsingS2D2}
\end{figure}

We report the result for the von Neumann entropy and for the second R\'enyi entropy in figure \ref{RealIsingS2D2}.
We also show the trace distance and the Schatten distance for $n=2$.
It is clear from the figure that the qualitative features of the CFT are well captured by these spin-chain results (like the horizon and the rough shape),
but the quantitative agreement is scarce.
Indeed, it is not even clear whether increasing $\ell$ the data are approaching the CFT predictions or not.
This fact has been already observed in Ref. \cite{Caputa:2016yzn} for the entanglement and R\'enyi entropies (as well as for the relative entropies):
even for values of $\ell$ of the order of few hundreds, deviations from the CFT have been observed
(see also \cite{ds-11,bertini-17,ep-07,ekpp-08,cc-13,irt-14,vs-17} for similar issues in other local quenches).
These discrepancies have a double origin: on the one hand, they are intrinsically connected with the finite lattice spacing, on the other, they  come from the
non-linearity of the dispersion relation \eqref{eps_IS}.
%
Hence, a natural way to circumvent at least the second problem and to provide more accurate tests for the CFT predictions is to linearize by hand the dispersion relation.
In this approach, we always use the Hamiltonian in Eq. \eqref{DH}, but, instead of having the actual dispersion of the critical Ising model \eqref{eps_IS}, we fix it by hand to
\be \label{vek1}
\ve_k = \f{2\pi|k|}{L}.
\ee
We denote the spin chain with this dispersion relation as the linearized critical Ising spin chain.
Notice that this dispersion is very artificial and the corresponding spin chain would have non-local terms, but all this is unimportant for our aims.
In this linearized critical Ising spin chain with the dispersion (\ref{vek1}), the integrals for the elements of the correlation matrix \eqref{CMfhg2} can be performed
analytically, obtaining
\bea \label{CMfhg3}
&& f_{j_1j_2} = - \f{2 \ii t^2 \sin (\pi t)}{\pi^2}
  \frac{(-)^{{j_1}-{\ell_0}} - (-)^{{j_2}-{\ell_0}}}
    {[({j_1}-{\ell_0})^2-t^2] [({j_2}-{\ell_0})^2-t^2]}, \nn\\
&& h_{j_1j_2} = \f{2 \ii t \cos (\pi t)}{\pi ^2}
  \frac{(-)^{{j_1}-{\ell_0}} ({j_2}-{\ell_0}+\frac{1}{2}) - (-)^{{j_2}-{\ell_0}}({j_1}-{\ell_0}+\frac{1}{2}) }
    {[({j_1}-{\ell_0}+\frac{1}{2})^2-t^2] [({j_2}-{\ell_0}+\frac{1}{2})^2-t^2]},\\
&& g_{j_1j_2} = - \f{\ii}{\pi} \f{1}{j_2-j_1+\f12}
  -\f{2 \ii t}{\pi ^2}
  \frac{(-)^{{j_1}-{\ell_0}} ({j_2}-{\ell_0}+\frac{1}{2}) \sin (\pi t) + (-)^{{j_2}-{\ell_0}}t \cos (\pi t) - (-)^{{j_1}+{j_2}-2 {\ell_0}} t}
    {[({j_1}-{\ell_0})^2-t^2] [({j_2}-{\ell_0}+\frac{1}{2})^2-t^2]}. \nn
\eea

The numerical results for entanglement entropy, R\'enyi entropy, trace distance, and Schatten distance in the linearized critical Ising spin chain are
shown in Figure \ref{LinearizedIsingS2D2}.
Although there are still some small deviations due to the intrinsic presence of the lattice spacing, it is clear that the data are converging very quickly to the CFT predictions
(cf. Eqs. (\ref{DSAnt}), (\ref{DSAt}) and (\ref{DAnt1t2Dt1t2}) with $q=1/2$) as $\ell$ becomes moderately large, especially when compared to the results for the
actual spin chain reported in Figure \ref{RealIsingS2D2}.
These findings confirm that most of the deviations observed for the actual spin chain in Figure \ref{RealIsingS2D2} come from the non-linearity of the dispersion relation,
i.e. must be imputed to the fact that the insertion of the local operator excites modes in the non-linear part of the spectrum of the lattice model.

\begin{figure}[p]
 \centering
 \includegraphics[height=1.2\textwidth]{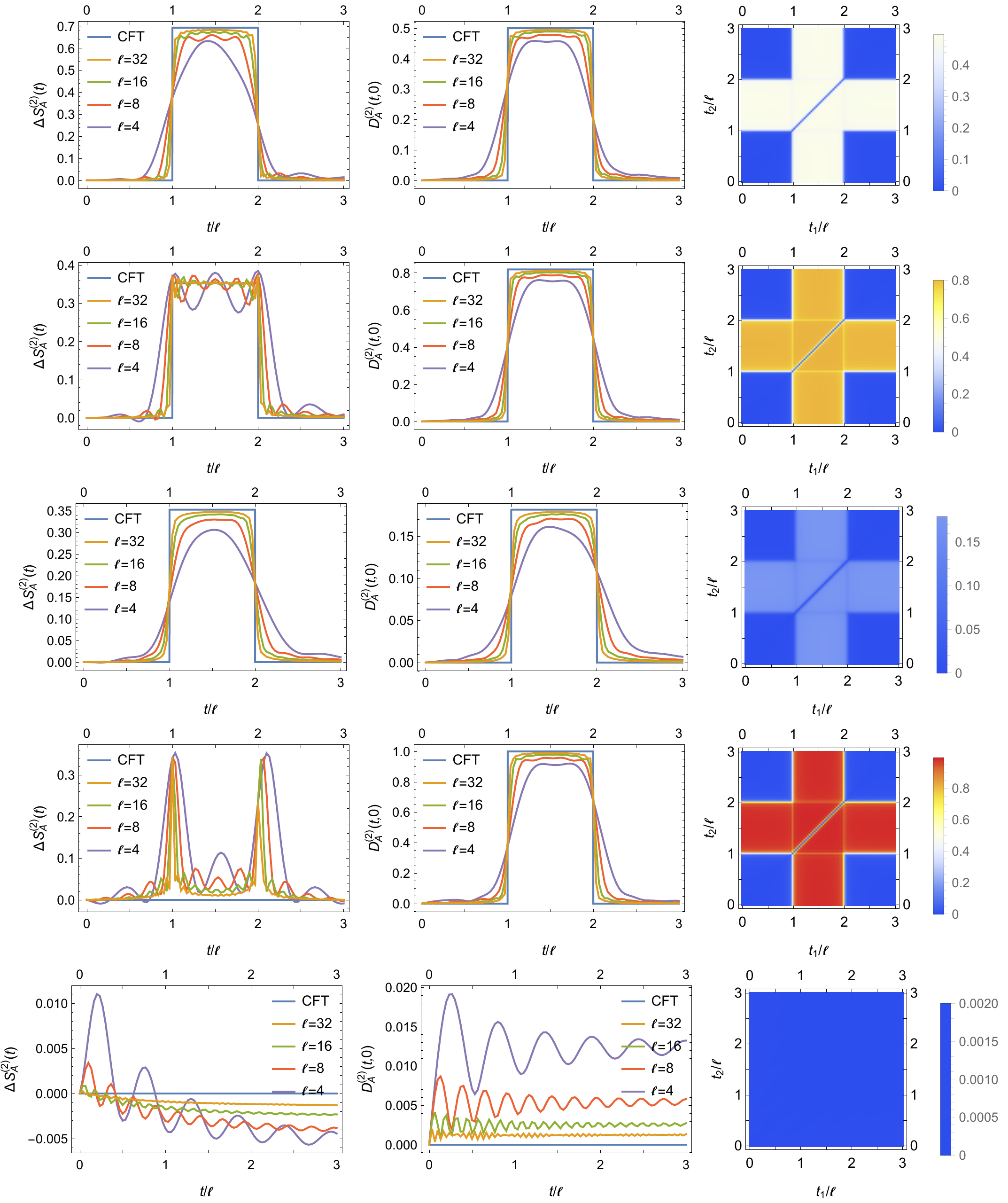}\\
 \caption{Local operator quench in the linearized critical Ising spin chain. 
 From top to bottom, we consider the states locally excited by $d_{2j}$, $d_{2j-1}+d_{2j}$, $d_{2j-1}-d_{2j}$, $\psi(j-\tf14)$, $\bar\psi(j+\tf14)$.
 For these states we have respectively ${q} = \f12, \f12-\f1\pi, \f12+\f1\pi,0,1$ (see Table \ref{tabIsing}).
 We focus on second R\'enyi entropy and Schatten distance.
 In the 3rd column we report the color plots of the Schatten 2-distance between RDM at generic times $D_A(t_1,t_2)$ with $\ell=32$. }
\label{LinearizedIsingS2D2X}
\end{figure}

In a similar way, we can investigate local operator quenches starting from other initial states in the critical Ising spin chain.
In Figure~\ref{LinearizedIsingS2D2X} we report five more examples of these quenches.
We only report the results for the second R\'enyi entropy and the Schatten distance of order 2 because we can access larger values of $\ell$ (compared to
the trace distance).
We limit ourselves to plot the numerical results for the linearized Ising chain since those with the original dispersion are not very illuminating.
For all the states we report (details in the caption of the Figure) the identification of the CFT operator corresponding to the lattice one is
not straightforward. In appendix \ref{appIde} we show how this identification is done.
As reported there, many simple states correspond to rather complicated values of $q$ (even non-rational), allowing us to test non-trivial
aspects of the CFT predictions.
From the Figure, it is evident that in all cases the numerical results quickly approach (as $\ell$ increases) the CFT predictions, also in those
cases when $q$ is a complicated number (e.g. for $q=1/2\pm 1/\pi$).
Consequently, all these results represent robust tests not only of the CFT predictions, but also of the correspondence between lattice and continuum operators
reported in Appendix \ref{appIde}.

Finally, we notice the resemblance of some of the curves in Figures~\ref{LinearizedIsingS2D2} and \ref{LinearizedIsingS2D2X} with the results for the
CFT with a finite cutoff ($\epsilon=1$) reported in Figure \ref{CFTfinitecutoff}.
This resemblance confirms that the remaining deviations in the linearized model are mainly due to the UV cutoff (i.e., the lattice spacing).
From the figure, it is clear that in a few cases (e.g. the fourth and the fifth rows) the approach to the CFT results is very peculiar and non-uniform.
In these cases, we did not work out the CFT prediction at finite $\e$ and hence we cannot identify the source of these deviations.

\subsection{The XX spin chain}

We now move our attention to the XX spin chain in zero field, i.e. Eq. \eqref{HXY} with $\g=\lambda=0$.
The construction of the RDM, as well as the derivation of entropies and distances, is identical to that of the Ising chain and we do not repeat it here.
We first considered the numerical calculations for the actual spin chain, but as for the Ising case, the results are affected by very large corrections
(they closely resemble those in Figure \ref{RealIsingS2D2}) and we do not find instructive or illuminating to report them here.
We then consider a linearized model. In this case, the Hamiltonian is always the one in Eq. \eqref{DH}, but with linear dispersion
\be \label{vek2}
\ve_k = \Big| \f{2\pi |k| }{L} - \f{\pi}{2} \Big|,
\ee
which we will call the linearized XX spin chain.
Notice the presence of two chiral linear modes, showing that, in this case, we deal with a Dirac fermion rather than a Majorana one (and hence $c=1$).

We consider many local operator quenches. The correspondence between lattice and CFT operators can be found again in Appendix \ref{appIde}.
In Figure~\ref{LinearizedXXS2D2X} we only report the numerical results for five of these initial states chosen as those with the most distinctive features
among those we considered.
For all the considered initial states, we have that increasing $\ell$ the results quickly approach the CFT predictions, similarly to what happens
for the linearized Ising chain in Figure \ref{LinearizedIsingS2D2X}.
Even in this case, we report a few results also for non-simple values of $q$ (e.g. $q=1/3,2/3$).

\begin{figure}[p]
 \centering
 \includegraphics[height=1.2\textwidth]{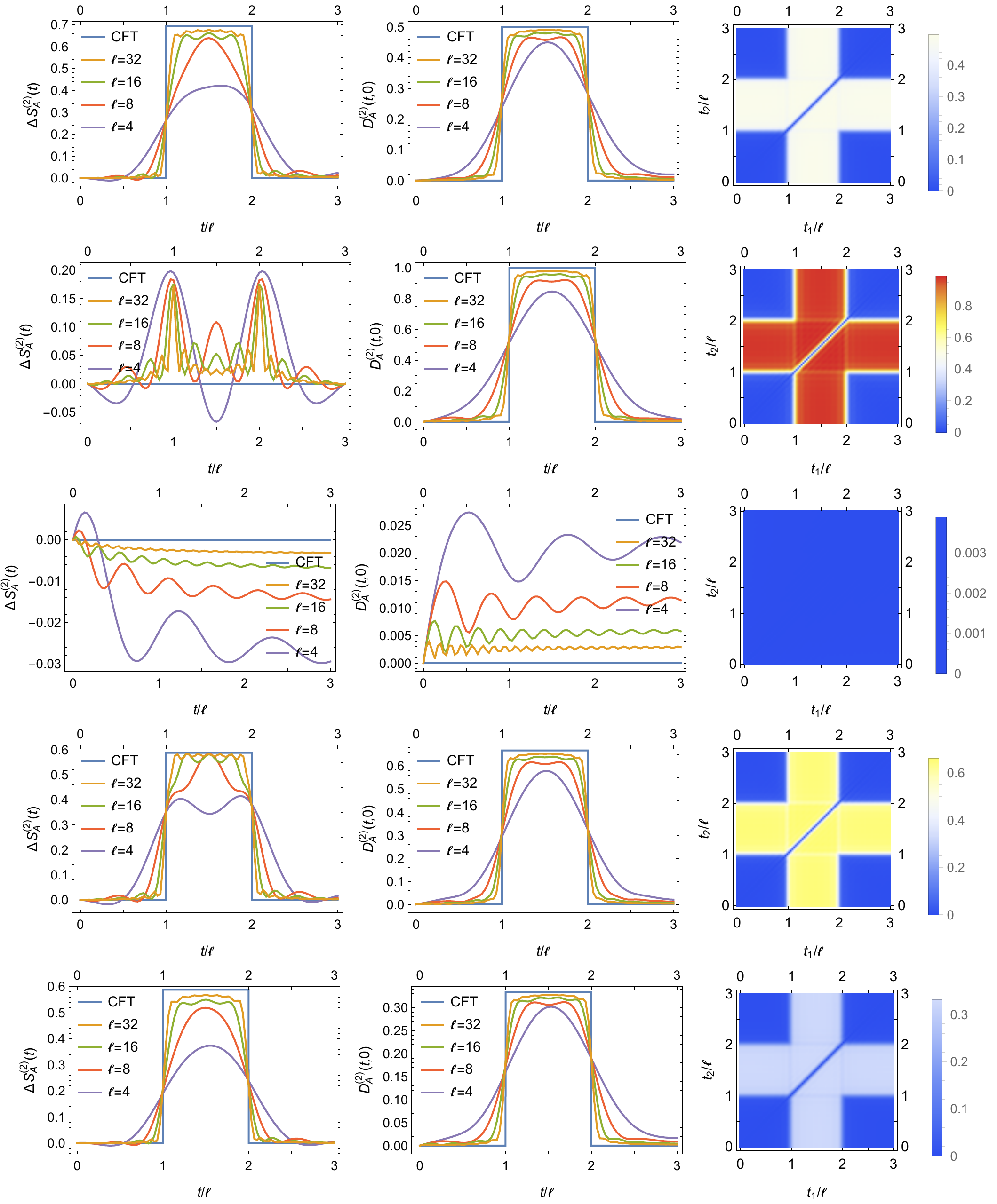}\\
 \caption{
 Local operator quench in the linearized XX spin chain in zero field. 
 From top to bottom, we consider the states locally excited by $a_j$, $\ep^{\ii\phi(j)}$, $\ep^{\ii\bar\phi(j)}$, $\ep^{\ii\phi(j)} + \ep^{-\ii\bar\phi(j)} + \ep^{-\ii\phi(j)}$,
 $ \ii ( \ep^{-\ii\bar\phi(j)} + \ep^{-\ii\phi(j)} - \ep^{\ii\bar\phi(j)} )$ with respectively ${q} = \f12,0,1,\f13,\f23$ (see Table \ref{tabXX}).
 We focus on second R\'enyi entropy and Schatten distance.
 In the 3rd column we report the color plots of the Schatten 2-distance between RDM at generic times $D_A(t_1,t_2)$ with $\ell=32$.
 }\label{LinearizedXXS2D2X}
\end{figure}

Note that in both Figure~\ref{LinearizedIsingS2D2X} for linearized Ising chain and Figure~\ref{LinearizedXXS2D2X} for linearized XX chain there are clear oscillations
in the second R\'enyi entropies and Schatten 2-distances.
This is a well known lattice artefact that directly follows from the oscillations of the correlation matrices and vanishes in the continuum limit (as clear from the figures).

\section{Operation insertion inside the interval}
\label{secInside}

Several generalizations in CFT of our results are completely straightforward. For example, if we start from an insertion of the operator on the right of the interval,
the results are identical by parity (i.e. we should switch holomorphic and anti-holomorphic features).
Another trivial generalization concerns finite systems for which the quasiparticle picture just leads to a periodic repetition of the time evolution we have found.
Similarly for systems with boundaries: one should only take care of the reflection of the quasiparticles at the edges.
Also the mutual information can be deduced from the combination of entanglement entropies, as well as more complicated entanglement related
quantities having a quasiparticle interpretation in CFT, as e.g. the negativity \cite{ctc-14,ac-19} and the entanglement of operators \cite{d-17,adm-19}.


A less trivial, but still straightforward generalization of our calculations is to consider
the case in which the insertion of the operator is performed inside the interval $A$.
The needed algebra is very similar to the one above; hence we just sketch the main steps of the derivation and we report the final results.
Without loss of generality, we choose the interval $A=[0,3\ell]$ with the general operator (\ref{cOcPcQcR}) inserted at the position $x=\ell$ at time $t=0$
(all other geometries can be easily deduced thanks to the quasiparticle picture).

Let us first discuss the entropies.
According to the quasiparticle picture, as long as the two particles produced by the insertion are still
within the interval, there is no increase of R\'enyi entropies. When the first quasiparticle passes through one of the edges,
there is a jump in the entropy (of the amount in Eq. \eqref{DS1}) which jumps back to its initial value when also the other
quasiparticle moves out of the interval.
Accidentally, with our choice of the relative distances between operator insertion and interval edges,
the R\'enyi and entanglement entropies have exactly the same functional form as in Eqs. \eqref{DS1} and \eqref{DSS}, i.e.
\be
\D S_A^{(n)}(t) =
\lt\{\ba{ll}
 0 & 0<t<\ell {~\rm and~} t>2\ell \\
 -\f{1}{n-1} \log \Big( p^n + q^n + \sum_i \f{r_i^n}{d_{\cR_i}^{n-1}} \Big) & \ell<t<2\ell
\ea\rt.,
\ee
\be
\D S_A(t) =
\lt\{\ba{ll}
 0                              & 0<t<\ell {~\rm and~} t>2\ell \\
 -p\log p - q \log q - \sum_i r_i \log\f{r_i}{d_{\cR_i}} & \ell<t<2\ell
\ea\rt..
\ee
We stress that this is only due to the fact that the left-moving quasiparticle exits the interval at time $t=\ell$ and the right moving one at
$t=2\ell$; in the general case, the threshold times are  given by the two separations between the edges and the point of operator insertion.

\begin{figure}[t]
  \centering
  \includegraphics[height=0.24\textwidth]{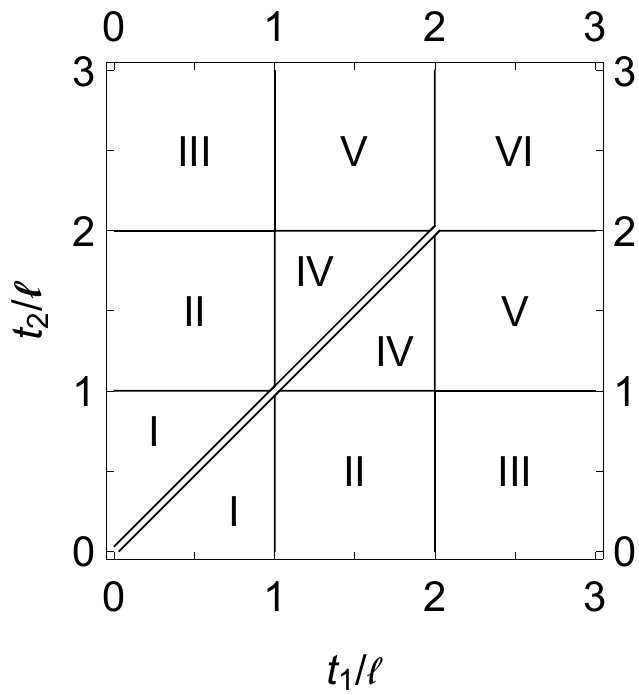}~~
  \includegraphics[height=0.24\textwidth]{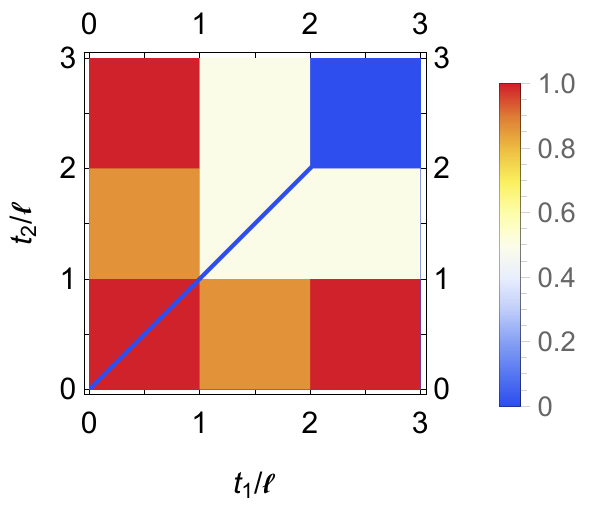}~~
  \includegraphics[height=0.24\textwidth]{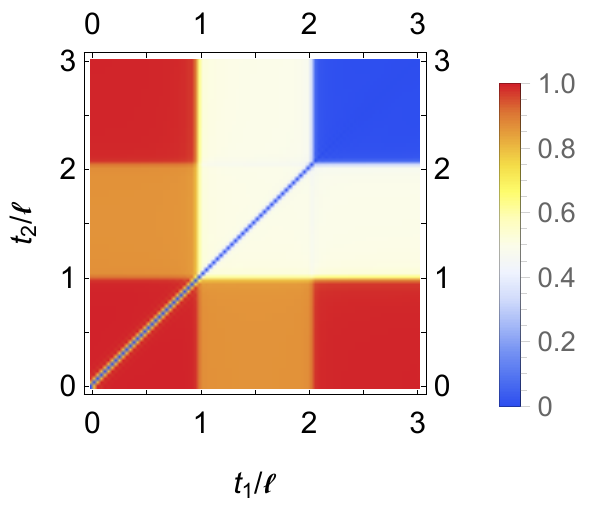}\\
  \caption{Local operator quench inside the subsystem $A=[0,3\ell]$ at $x=\ell$.
  Left: The division of the $(t_1,t_2)$ plane according to the positions of the quasiparticles.
  Middle: The CFT result of the second Schatten distance (\ref{insideDAnt1t2}) with $n=2,q=\f12$.
  Right: The numerical result of the second Schatten distance for the inserted operator $d_{2j-1}$ in the linearized critical Ising
  spin chain (with $\ell=32$, i.e. with an interval length $3\ell=96$).}
  \label{insideIsing}
\end{figure}

For the RDM distances, the situation is slightly more complicated. We can straightforwardly identify {\it six} regions in the $(t_1,t_2)$ plane
according to the relative positions of the quasiparticles. 
These regions are shown in the first panel of Figure \ref{insideIsing}.
At this point, one proceeds according to the same line of thoughts of Sec. \ref{sumpr}.
The main point is to use the orthogonality of RDMs with quasiparticles  at different positions.
For the Schatten distances, the final results are
\be
D_A^{(n)}(t_1,t_2)=
\lt\{\ba{ll}
 1                                                                                    & {\rm regions~I~and~III} \\
 \Big[ \f12 \Big( 1 + p^n + q^n + \sum_i \f{r_i^n}{d_{\cR_i}^{n-1}} \Big) \Big]^{1/n} & {\rm region~II} \\
 \Big( p^n + \sum_i \f{r_i^n}{d_{\cR_i}^{n-1}} \Big)^{1/n}                            & {\rm region~IV} \\
 \Big[ \f12 \Big( p^n + (1-q)^n + \sum_i \f{r_i^n}{d_{\cR_i}^{n-1}} \Big) \Big]^{1/n} & {\rm region~V} \\
 0                                                                                    & {\rm region~VI}
\ea\rt.,
\label{Dninside}
\ee
leading, by replica limit $n\to1$,  to the trace distance
\be \label{insideDAt1t2}
D_A(t_1,t_2)=
\lt\{\ba{ll}
 1   & {\rm regions~I,~II~and~III} \\
 1-q & {\rm regions~IV~and~V} \\
 0 & {\rm region~VI} \\
\ea\rt..
\ee
It is remarkable that, once again, the trace distance only depends on the ratio of the purely anti-holomorphic part $q$, in spite of the
more complex results for the Schatten distances in Eq. \eqref{Dninside}.

\begin{figure}[t]
  \centering
  \includegraphics[width=\textwidth]{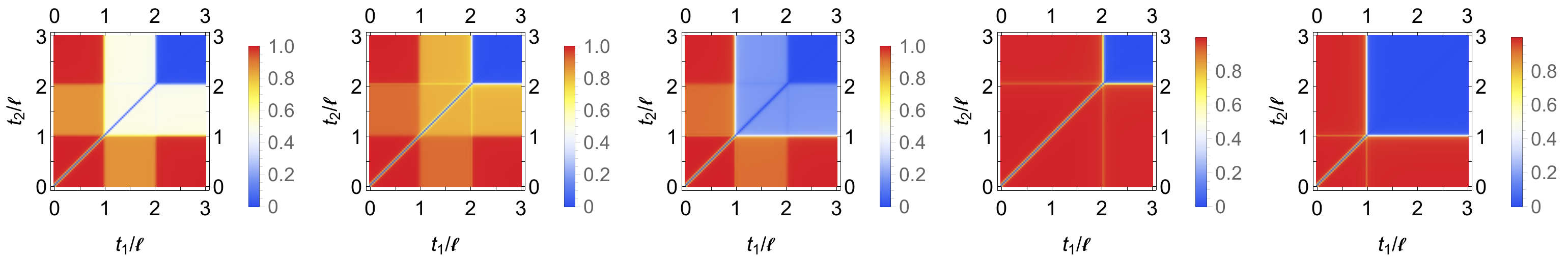}\\
  \includegraphics[width=\textwidth]{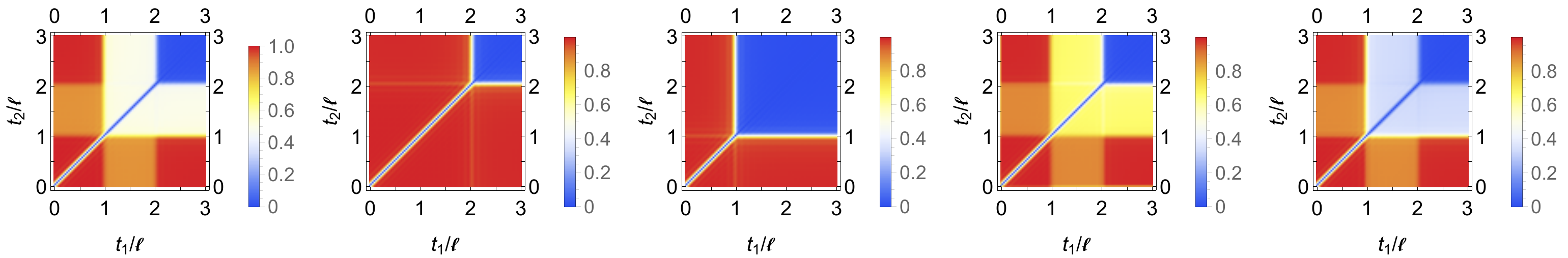}\\
  \caption{The second Schatten distance after local operator quench inside the subsystem $A=[1,3\ell=96]$ in spin chains.
  The operator is always inserted at $\ell=32$.
  Top: Linearized critical Ising chain. From left to right the inserted operators are $d_{2j}$, $d_{2j-1}+d_{2j}$, $d_{2j-1}-d_{2j}$, $\psi(j-\tf14)$, $\bar\psi(j+\tf14)$, respectively, and the corresponding ratios of the purely anti-holomorphic parts are ${q} = \f12, \f12-\f1\pi, \f12+\f1\pi,0,1$.
  Bottom: Linearized critical XX chain in zero field. From left to right the inserted operators are $a_j$, $\ep^{\ii\phi(j)}$, $\ep^{\ii\bar\phi(j)}$, $\ep^{\ii\phi(j)} + \ep^{-\ii\bar\phi(j)} + \ep^{-\ii\phi(j)}$, $\ii(\ep^{-\ii\bar\phi(j)} + \ep^{-\ii\phi(j)} - \ep^{\ii\bar\phi(j)})$ with respectively ${q} = \f12,0,1,\f13,\f23$.
  These numerical results perfectly match the CFT prediction (\ref{insideDAnt1t2}).}
  \label{insideIsingXXXx}
\end{figure}

Specializing to the case of an the inserted operator being the sum of a purely holomorphic and a purely anti-holomorphic primary operators
(i.e. the case with $p+q=1$ and $r_i=0$), we get the Schatten distance
\be \label{insideDAnt1t2}
D_A^{(n)}(t_1,t_2)=
\lt\{\ba{ll}
 1                                                    & {\rm regions~I~and~III} \\
 \Big[ \f12 \Big( 1 + (1-q)^n + q^n \Big) \Big]^{1/n} & {\rm region~II} \\
 1-q                                                  & {\rm regions~IV~and~V} \\
 0                                                    & {\rm region~VI}
\ea\rt..
\ee
Notice that there are four different values in the Schatten distance (\ref{insideDAnt1t2}), while there are three different values in the trace distance (\ref{insideDAt1t2}).
In Figures ~\ref{insideIsing} and \ref{insideIsingXXXx} we report the CFT results for the second Schatten distance for several operator insertions for both
the fermion and boson theory.
The CFT predictions are tested against numerical results in the critical Ising and XX {\it linearized} spin chains (with $\ell=32$), finding perfect agreement.
We used the correspondence between lattice and continuum operators reported in Appendix \ref{appIde}.

\section{Conclusions and discussions}\label{secCon}

We have investigated the time evolution of entanglement entropy, R\'enyi entropy, trace distance, and Schatten distance for an interval embedded in the infinite line after
local operator quenches in 2D CFT and one-dimensional critical Ising and XX spin chains.
We obtained analytical results in the CFT that match the numerical results in the spin chain with a linearized dispersion relation.
All our results are consistent with the picture of entangled quasiparticles put forward in \cite{Calabrese:2005in}.
In the setting of local operator quench, a nonchiral primary operator excites both left-moving and right-moving quasiparticles,
a holomorphic primary operator only excites a right-moving quasiparticle and an anti-holomorphic primary operator only excites a left-moving one.
The RDM of the interval hosting a quasiparticle is orthogonal to the RDM of the interval without any quasiparticles.
Moreover, the RDMs of two intervals hosting quasiparticles at different positions are also orthogonal to each other.
For a mixed operator (i.e. for the sum of primaries with different chiralities), it is remarkable that all the quantities we investigated only depend on the
ratio of the anti-holomorphic particles.

Our results show that subsystem distances for a local operator quench provide information that the entanglement entropy does not.
For example, the entanglement entropy cannot distinguish the quench initiated by a purely holomorphic operator or by a purely anti-holomorphic operator, while the distance does.
Another example, concerns the quench initiated by a purely holomorphic operator.
The entanglement entropy would provide the same results in the following three cases:
(1) no quasiparticles are excited,
(2) both left-moving and right-moving quasiparticles are excited, but they are not entangled, and
(3) only a right-moving quasiparticle is excited.
Thus, the entanglement entropy cannot distinguish between the three while the subsystem distances identify (3) as the correct answer.
More examples of this sort are very easily constructed.

We have seen that the agreement of the CFT predictions with the data of the true spin chains is quite scarce if we do not linearize the dispersion relation.
It would be interesting, for the spin chains considered here (and more generically for an arbitrary integrable system), to find a procedure to take into account the contributions of the
various excited modes with different velocities, for example on the lines of what has been done for the entanglement entropy after a global quantum
quench \cite{alba-2016,alba-2018}.

In CFT, we have investigated the insertion of a general operator (\ref{cOcPcQcR}), while in spin chains we have only checked the cases with special operators of the type (\ref{cOwbarw}). As there is one-to-one correspondence between the low energy operators in the CFT and spin chain, it would be interesting to test also other cases.
The recent progresses on the identification of CFT and spin chain local operators in \cite{Zou:2019dnc,Zou:2019iwr} could help in this process.

In this paper, we focused on free boson the fermion theories, but it would be interesting to investigate the subsystem distances in holographic CFTs, which are CFTs
with classical gravity duals \cite{Maldacena:1997re}.
Recently, a new method of entanglement wedge reconstruction using distinguishability measure of RDMs of locally excited states has been proposed \cite{Suzuki:2019xdq,Kusuki:2019hcg}.
The considered locally excited states are similar to the ones in this paper.
Hence, as already proposed \cite{Suzuki:2019xdq,Kusuki:2019hcg}, it is surely worth to generalize their calculations to the subsystem trace distance.

\section*{Acknowledgments}

We thank Paola Ruggiero for helpful discussions.
Both authors acknowledge support from ERC under Consolidator grant number 771536 (NEMO).

\appendix

\section{Identification of CFT and spin chain operators}\label{appIde}

In this appendix, we establish the correspondence between the inserted local operator in the CFT and in the spin chain.
To this aim, we first write the CFT time evolved state and then match it with the spin chain state in the continuum limit.
This allows deriving the ratio $q$ of the anti-holomorphic part (\ref{ratio}), which is the only quantity entering in the various formulas
for entropies and distances.

\subsection{Time evolved state in CFT}

For a 2D CFT on a cylinder with spatial period $L$ we use the coordinates
\be
w= \t-\ii x= \ii(t-x), ~~ \bar w= \t+\ii x= \ii(t+x).
\ee
Here $\t$ is the Euclidean time, $t$ is the real time, and $x$ the periodic spatial coordinate $x= x+L$.
The cylinder can be mapped to the complex plane $z$ by
\be
z= \ep^{\f{2\pi w}{L}} = \ep^{\f{2\pi\ii}{L}(t-x)},~~
\bar z= \ep^{\f{2\pi \bar w}{L}} = \ep^{\f{2\pi\ii}{L}(t+x)}.
\ee
For a primary operator $\cO$ with conformal weights $(h_\cO,\bar h_\cO)$, i.e. scaling dimension $\D_\cO=h_\cO+\bar h_\cO$ and spin $s_\cO=h_\cO-\bar h_\cO$, we have
\be
\cO(t,x) = \cO(w,\bar w) = \Big(\f{\p z}{\p w}\Big)^{h_\cO} \Big(\f{\p \bar z}{\p \bar w}\Big)^{\bar h_\cO} \cO(z,\bar z).
\ee
On the complex plane, the primary operator can be expanded as
\be
\cO(z,\bar z) = \sum_{m,\bar m \in \rZ} z^m\bar z^{\bar m} \cO_{-h_\cO-m,-\bar h_\cO-\bar m},
\ee
with modes satisfying
\be
\cO_{-h_\cO-m,-\bar h_\cO-\bar m} |0\rag = 0, ~ m<0 ~{\rm or}~ \bar m<0.
\ee
For $m\geq0$ and $\bar m\geq0$ we also have
\be
\cO_{-h_\cO-m,-\bar h_\cO-\bar m} |0\rag = \sr{\a_{h_\cO,\bar h_\cO}^{m,\bar m}} |\cO_{-h_\cO-m,-\bar h_\cO-\bar m}\rag,
\ee
with the normalization
\be
\a_{h_\cO,\bar h_\cO}^{m,\bar m} = \lag 0|\cO^\dag_{h_\cO+m,\bar h_\cO+\bar m} \cO_{-h_\cO-m,-\bar h_\cO-\bar m} |0\rag = \f{(2h_\cO+m-1)!(2\bar h_\cO+\bar m-1)!}{m!\bar m!(2h_\cO-1)!(2\bar h_\cO-1)!},
\ee
($\cO$ has been normalized so that $\a=1$ for $m=\bar m=0$)
and the orthonormality condition
\be
\lag \cO_{-h_\cO-m_1,-\bar h_\cO-\bar m_1} |\cO_{-h_\cO-m_2,-\bar h_\cO-\bar m_2}\rag = \d_{m_1m_2}\d_{\bar m_1\bar m_2}.
\ee

We are now ready to write down the state with the insertion of a local primary operator \cite{Zou:2019dnc}
\be
\cO(t,x) |0\rag = \Big( \f{2\pi}{L} \Big)^{\D_\cO} \sum_{m,\bar m=0}^{+\inf} \ep^{ \f{2\pi\ii}{L}[ (h_\cO+m)(t-x) + (\bar h_\cO+\bar m)(t+x) ]} \sr{\a_{h_\cO,\bar h_\cO}^{m,\bar m}} |\cO_{-h_\cO-m,-\bar h_\cO-\bar m}\rag,
\ee
which leads to
\be
\ep^{-\ii H t} \cO(0,x) |0\rag = \Big( \f{2\pi}{L} \Big)^{\D_\cO} \ep^{\f{\pi\ii ct}{6L}}
\sum_{m,\bar m=0}^{+\inf} \ep^{-\f{2\pi\ii}{L}[ (h_\cO+m)(t+x) + (\bar h_\cO+\bar m)(t-x) ]} \sr{\a_{h_\cO,\bar h_\cO}^{m,\bar m}} |\cO_{-h_\cO-m,-\bar h_\cO-\bar m}\rag.
\ee
Here $c$ is the central charge of the 2D CFT.
Specializing to a purely holomorphic primary operator $\cP(t,x)=\cP(w)$ with conformal weights $(\f12,0)$ and to a purely anti-holomorphic primary operator
$\cQ(t,x)=\cQ(\bar w)$ with conformal weights $(0,\f12)$ we have
\bea \label{cPtx0barxPtx0}
&& \ep^{-\ii H t} \cP(0,x) |0\rag = \sr{\f{2\pi}{L}} \ep^{\f{\pi\ii ct}{6L}} \sum_{k=0}^{+\inf} \ep^{-\f{2\pi\ii}{L}(\f12+k)(t+x)}|\cP_{-\f12-k}\rag , \nn\\
&& \ep^{-\ii H t} \cQ(0,x) |0\rag = \sr{\f{2\pi}{L}} \ep^{\f{\pi\ii ct}{6L}} \sum_{k=0}^{+\inf} \ep^{-\f{2\pi\ii}{L}(\f12+k)(t-x)}|\cQ_{-\f12-k}\rag .
\eea
Keep in mind that $|\cP_{-\f12-k}\rag$ and $|\cQ_{-\f12-k}\rag$ are orthonormal states.
Note also that the two states $\ep^{-\ii H t} \cP(0,x) |0\rag$  and $\ep^{-\ii H t} \cQ(0,x) |0\rag$ in Eq. (\ref{cPtx0barxPtx0}) are not regularized nor normalized.

\subsection{Operator correspondence in the critical Ising spin chain}

In XY spin chain, the insertion of the Majorana modes (\ref{d2jm1d2j}) in the ground state leads to the time-dependent states
\bea
&& \ep^{-\ii H t} d_{2j-1} | 0 \rag = \f{1}{\sr{L}} \sum_k \ep^{\ii( j\vph_k + \f{\th_k}{2} - \ve_k t )} c_k^\dag |0\rag , \nn\\
&& \ep^{-\ii H t} d_{2j} | 0 \rag = -\f{\ii}{\sr{L}} \sum_k \ep^{\ii( j\vph_k - \f{\th_k}{2} - \ve_k t )} c_k^\dag |0\rag .
\eea
In the critical Ising spin chain with $\g=\l=1$, the corresponding CFT is the 2D free massless fermion theory.
We use the linearized dispersion relation (\ref{vek1}) and get
\bea \label{epmiiHtd2jm1epmiiHtd2j}
&& \ep^{-\ii H t} d_{2j-1} | 0 \rag = \f{1}{\sr{L}} \sum_{k=0}^{\f{L}{2}-1}
  \big( \ep^{-\f{2\pi\ii}{L}(\f12+k)(t+j-\f14)} \ep^{-\f{\pi\ii}{4}}c^\dag_{-\f12-k}|0\rag
  + \ep^{-\f{2\pi\ii}{L}(\f12+k)(t-j+\f14)} \ep^{\f{\pi\ii}{4}}c^\dag_{\f12+k}|0\rag \big), \nn\\
&& \ep^{-\ii H t} d_{2j} | 0 \rag = \f{1}{\sr{L}} \sum_{k=0}^{\f{L}{2}-1}
  \big( \ep^{-\f{2\pi\ii}{L}(\f12+k)(t+j+\f14)} \ep^{-\f{\pi\ii}{4}}c^\dag_{-\f12-k}|0\rag
  - \ep^{-\f{2\pi\ii}{L}(\f12+k)(t-j-\f14)} \ep^{\f{\pi\ii}{4}}c^\dag_{\f12+k}|0\rag \big).
\eea
We compare the time-dependent CFT states (\ref{cPtx0barxPtx0}) with the spin chain states (\ref{epmiiHtd2jm1epmiiHtd2j}).
As the CFT states $\ep^{-\ii H t}\cP(0,x) |0\rag$ and $\ep^{-\ii H t}\cQ(0,x) |0\rag$ (\ref{cPtx0barxPtx0}) are not regularized and normalized,
we will not keep track of the overall normalization of the operators (the interested reader can find some recent progress on more careful identification of the CFT
and spin chain local operators in \cite{Zou:2019dnc,Zou:2019iwr}).
We find that in (\ref{epmiiHtd2jm1epmiiHtd2j}) the modes with negative momenta correspond to the holomorphic state $|\psi\rag$ and its derivatives,
and the modes with positive momenta correspond to the anti-holomorphic state $|\bar\psi\rag$ and its derivatives.
Up to the overall normalization, we identify
\be \label{idIsing}
d_{2j-1} = \psi(j-\tf14) + \bar\psi(j-\tf14), ~~
d_{2j} = \psi(j+\tf14) - \bar\psi(j+\tf14).
\ee
As all operators $d_{2j-1}$, $d_{2j}$, $\psi$ and $\bar\psi$ are Hermitian, in the above identification there are no phase ambiguities.
Also, the possible presence of minus signs can be absorbed in a redefinition of $\psi$ and $\bar\psi$.
The identifications (\ref{idIsing}) are consistent with recent results \cite{Zou:2019iwr}.
We get that for both $d_{2j-1}$ and $d_{2j}$ the ratio of anti-holomorphic part is ${q}=\f12$.
Since for $d_{2j-1}$ and $d_{2j}$ the corresponding CFT operators are inserted at slightly different positions, we should be careful when making combinations of the two.

\begin{table}[t]
 \centering
\begin{tabular}{|c|cccc|}\hline
 \small operator
 & $d_{2j-1},d_{2j}, a^{(\dag)}_j$ & $d_{2j-1}\pm d_{2j}$ &  $\psi(j \pm \tf14)$  & $\bar\psi(j\pm\tf14)$ \\
 \hline
 $q$  &  $\f12$ &   $\f12\mp\f1\pi$ & 0&1 \\
 \hline
\end{tabular}
 \caption{The locally excited states we consider in critical Ising spin chain.
 In the first row we give the operators that excite the states.
 In the second row ${q}$ is calculated as the ratio of the anti-holomorphic part, i.e. left-moving parts in our convention.}\label{tabIsing}
\end{table}

For $a_j=\f{1}{2}(d_{2j-1}-\ii d_{2j})$, we get the state
\bea
&& \ep^{-\ii H t} a_j | 0 \rag = \f{1}{2\sr{L}} \sum_{k=0}^{\f{L}{2}-1}
  \big[ ( \ep^{-\f{2\pi\ii}{L}(\f12+k)(t+j-\f14)} - \ii \ep^{-\f{2\pi\ii}{L}(\f12+k)(t+j+\f14)}) \ep^{-\f{\pi\ii}{4}}c^\dag_{-\f12-k}|0\rag \nn\\
&& \phantom{\ep^{-\ii H t} a_j | 0 \rag =}
  + ( \ep^{-\f{2\pi\ii}{L}(\f12+k)(t-j+\f14)} + \ii \ep^{-\f{2\pi\ii}{L}(\f12+k)(t-j-\f14)}) \ep^{\f{\pi\ii}{4}}c^\dag_{\f12+k}|0\rag \big],
\eea
from which we can read the ratio of the anti-holomorphic part ${q}=\f12$.
Similarly, for $a^\dag_j=\f{1}{2}(d_{2j-1}+\ii d_{2j})$, we get the state $\ep^{-\ii H t} a^\dag_j | 0 \rag$ with ${q}=\f12$.
We can also make the combination
\bea
&& \ep^{-\ii H t} ( d_{2j-1} + d_{2j} ) | 0 \rag = \f{1}{\sr{L}} \sum_{k=0}^{\f{L}{2}-1}
  \big[ ( \ep^{-\f{2\pi\ii}{L}(\f12+k)(t+j-\f14)} + \ep^{-\f{2\pi\ii}{L}(\f12+k)(t+j+\f14)}) \ep^{-\f{\pi\ii}{4}}c^\dag_{-\f12-k}|0\rag \nn\\
&& \phantom{\ep^{-\ii H t} a_j | 0 \rag =}
  + ( \ep^{-\f{2\pi\ii}{L}(\f12+k)(t-j+\f14)} - \ep^{-\f{2\pi\ii}{L}(\f12+k)(t-j-\f14)}) \ep^{\f{\pi\ii}{4}}c^\dag_{\f12+k}|0\rag \big],
\eea
for which the ratio of the anti-holomorphic part is
\be
{q} = \f{1}{L}\sum_{k=0}^{\f{L}{2}-1} \Big( 1 - \cos\f{\pi(k+\f12)}{L} \Big) = \f12 - \f{1}{2L\sin\f{\pi}{2L}} \to \f12 -\f1\pi ~{\rm as}~ L \to \inf.
\ee
Similarly, for the combination $\ep^{-\ii H t} ( d_{2j-1} - d_{2j} ) | 0 \rag$ we have
\be
{q} = \f12 + \f{1}{2L\sin\f{\pi}{2L}} \to \f12 + \f1\pi ~{\rm as}~ L \to \inf.
\ee
Furthermore, we can split each of the states in (\ref{epmiiHtd2jm1epmiiHtd2j}) into two parts as
\bea
&& \ep^{-\ii H t} d_{2j-1} | 0 \rag = \ep^{-\ii H t} \psi(j-\tf14) | 0 \rag + \ep^{-\ii H t} \bar\psi(j-\tf14) | 0 \rag , \nn\\
&& \ep^{-\ii H t} d_{2j} | 0 \rag = \ep^{-\ii H t} \psi(j+\tf14) | 0 \rag - \ep^{-\ii H t} \bar\psi(j+\tf14) | 0 \rag.
\eea
These four operators $\psi(j\mp\tf14)$ and $\bar\psi(j\mp\tf14)$ are not local operators in the spin chain, but they correspond to (quasi)-local operators in the CFT
and the corresponding states are also locally excited states.
It is easy to see that $\ep^{-\ii H t} \psi(j-\tf14) | 0 \rag$ and $\ep^{-\ii H t} \psi(j+\tf14) | 0 \rag$ have ${q}=0$ and $\ep^{-\ii H t} \bar\psi(j-\tf14) | 0 \rag$ and $\ep^{-\ii H t} \bar\psi(j-\tf14) | 0 \rag$ have ${q}=1$.

We summarize the value of $q$ for all the excited states we consider in the critical Ising spin in table~\ref{tabIsing}.


\subsection{Operator correspondence in the XX spin chain}

 \begin{table}[t]
 \centering
\begin{tabular}{|c|cccc|}\hline
 \small operator
 & $d_{2j-1}, d_{2j} ,a^{(\dag)}_j$ &$d_{2j-1}\pm d_{2j}$ &   $\ep^{\pm \ii\phi(j)}$, $\cos(\phi(j))$  & $\ep^{\pm\ii\bar\phi(j)}$, $ \sin(\bar\phi(j))$ \\
 \hline
 $q$  &  $\f12$ & $\f12$  & 0 & 1 \\
 \hline
 \hline
  \small operator
 & $\ep^{\ii\phi(j)}+\ep^{\ii\bar\phi(j)}$ & $-\ii ( \ep^{\ii\phi(j)} - \ep^{-\ii\bar\phi(j)} )$ & $\ep^{\ii\phi(j)} + \ep^{-\ii\bar\phi(j)} + \ep^{-\ii\phi(j)}$ & $ \ii ( \ep^{-\ii\bar\phi(j)} + \ep^{-\ii\phi(j)} - \ep^{\ii\bar\phi(j)} )$
 \\
 \hline
 $q$  &  $\f12$ & $\f12$  & $\f13$ & $\f23$ \\
 \hline
\end{tabular}
 \caption{Some of the locally excited states we consider in critical XX spin chain.
 We give the  operators that excite the states and the corresponding ratio of the anti-holomorphic part ${q}$.
 Note that there are more states we have not included in the table. }\label{tabXX}
\end{table}

In the critical XX spin chain with $\g=\l=0$, we proceed exactly in the same way as above.
Using the linearized dispersion relation (\ref{vek2}), we get
\bea\label{epmiiHtd2jm1epmiiHtd2jXX}
&& \ep^{-\ii H t}d_{2j-1}|0\rag = \f{1}{\sr{L}} \sum_{k=0}^{\f{L}{4}-1}
\big(
  \ep^{-\f{\pi\ii j}{2}} \ep^{-\f{2\pi\ii}{L}(\f12+k)(t+j)} c^\dag_{-\f{L}{4}-\f12-k} |0\rag
 + \ep^{-\f{\pi\ii (j+1)}{2}} \ep^{-\f{2\pi\ii}{L}(\f12+k)(t-j)} c^\dag_{-\f{L}{4}+\f12+k} |0\rag \nn\\
&& \phantom{\ep^{-\ii H t}d_{2j-1}|0\rag =}
 + \ep^{\f{\pi\ii (j+1)}{2}} \ep^{-\f{2\pi\ii}{L}(\f12+k)(t+j)} c^\dag_{\f{L}{4}-\f12-k} |0\rag
 + \ep^{\f{\pi\ii j}{2}} \ep^{-\f{2\pi\ii}{L}(\f12+k)(t-j)} c^\dag_{\f{L}{4}+\f12+k} |0\rag
\big), \nn\\
&& \ep^{-\ii H t}d_{2j}|0\rag = - \f{\ii}{\sr{L}} \sum_{k=0}^{\f{L}{4}-1}
\big(
  \ep^{-\f{\pi\ii j}{2}} \ep^{-\f{2\pi\ii}{L}(\f12+k)(t+j)} c^\dag_{-\f{L}{4}-\f12-k} |0\rag
 + \ep^{-\f{\pi\ii (j-1)}{2}} \ep^{-\f{2\pi\ii}{L}(\f12+k)(t-j)} c^\dag_{-\f{L}{4}+\f12+k} |0\rag \nn\\
&& \phantom{\ep^{-\ii H t}d_{2j-1}|0\rag =}
 + \ep^{\f{\pi\ii (j-1)}{2}} \ep^{-\f{2\pi\ii}{L}(\f12+k)(t+j)} c^\dag_{\f{L}{4}-\f12-k} |0\rag
 + \ep^{\f{\pi\ii j}{2}} \ep^{-\f{2\pi\ii}{L}(\f12+k)(t-j)} c^\dag_{\f{L}{4}+\f12+k} |0\rag
\big).
\eea
Comparing (\ref{cPtx0barxPtx0}) with (\ref{epmiiHtd2jm1epmiiHtd2jXX}) (without keeping track of the overall normalization), we get the tentative identification
\bea \label{idXX}
&& d_{2j-1} = \ep^{\ii\phi(j)} + \ep^{-\ii\bar\phi(j)} + \ep^{-\ii\phi(j)} + \ep^{\ii\bar\phi(j)}, \nn\\
&& d_{2j} = -\ii ( \ep^{\ii\phi(j)} - \ep^{-\ii\bar\phi(j)} - \ep^{-\ii\phi(j)} + \ep^{\ii\bar\phi(j)} ),
\eea
with $\phi$ and $\bar\phi$ being respectively the holomorphic and anti-holomorphic part of the scalar field.
More precisely, we have the identifications of the spin chain modes and CFT states
\bea
c^\dag_k|0\rag \with k = -\f{L}{2}+\f12,\cdots,-\f{L}{4}-\f12 &\lra& |\ep^{\ii\phi}\rag \rm{~and~its~derivatives}, \nn\\
c^\dag_k|0\rag \with k = -\f{L}{4}+\f12,\cdots,-\f12 &\lra& |\ep^{-\ii\bar\phi}\rag \rm{~and~its~derivatives}, \nn\\
c^\dag_k|0\rag \with k = \f12,\cdots,\f{L}{4}-\f12 &\lra& |\ep^{-\ii\phi}\rag \rm{~and~its~derivatives}, \nn\\
c^\dag_k|0\rag \with k = \f{L}{4}+\f12,\cdots,\f{L}{2}-\f12 &\lra& |\ep^{\ii\bar\phi}\rag \rm{~and~its~derivatives}.
\eea
In the identification (\ref{idXX}) there are phase ambiguities, but they do not affect the following calculations of $q$.
For all the states locally excited by $d_{2j-1}$, $d_{2j}$, $a_j$, $a^\dag_j$, $d_{2j-1}+d_{2j}$, $d_{2j-1}-d_{2j}$, we always have the ratio of the anti-holomorphic part ${q}=\f12$.
In the XX spin chain, we also have the states excited by $\ep^{\pm\ii\phi(j)}$ with ${q}=0$ and states $\ep^{\pm\ii\bar\phi(j)}$ with ${q}=1$.
Using the states excited by $\ep^{\pm\ii\phi(j)}$, $\ep^{\pm\ii\bar\phi(j)}$, we can construct more states in spin chain that correspond to locally excited state in CFT.
For examples, we have a state excited by $\cos\phi(j)=\f12( \ep^{\ii\phi(j)} + \ep^{-\ii\phi(j)} )$ with ${q}=0$, a state excited by $\ep^{\ii\phi(j)} + \ep^{-\ii\bar\phi(j)} + \ep^{-\ii\phi(j)}$ with ${q}=\f13$, and a state excited by $-\ii (- \ep^{-\ii\bar\phi(j)} - \ep^{-\ii\phi(j)} + \ep^{\ii\bar\phi(j)} )$ with ${q}=\f23$.

We summarize some of the states we have considered in table~\ref{tabXX}.
Note that there are more states we have not included in the table.


\providecommand{\href}[2]{#2}\begingroup\raggedright\endgroup


\end{document}